\documentclass[12pt]{article}
\usepackage[margin=1in]{geometry}
\usepackage{amsmath,amsfonts,amssymb,amsthm,epsfig, graphicx, hyperref, mathtools}
\usepackage[dvipsnames,table,dvipsnames*, svgnames*, hyperref]{xcolor}
\usepackage{natbib}
\usepackage{setspace}
\usepackage{enumitem}
\usepackage{float}

\numberwithin{equation}{section}
\theoremstyle{plain}
\newtheorem{theorem}{Theorem}

\theoremstyle{definition}

\newcommand{\ignore}[1]{}
\newcommand{\beq}{\begin{equation}}
\newcommand{\eeq}{\end{equation}}
\newcommand{\bet}{\begin{theorem}}
\newcommand{\eet}{\end{theorem}}
\newcommand{\R}{\mathbb{R}}
\newcommand{\E}{\mathbb{E}}
\title{Inference of Microbial Interactions Using Copula Models with Mixture Margins} 
\author{Rebecca A. Deek and Hongzhe Li}
\date{Department of Biostatistics, Epidemiology, and Informatics, Perelman School of Medicine, University of Pennsylvania, Philadelphia, PA, 19104}

\begin{document}

\maketitle

\begin{abstract}
	Quantification of microbial interactions from 16S rRNA and meta-genomic sequencing data is difficult due to their sparse nature, as well as the fact that the data only provides measures of relative abundance. In this paper, we propose using copula models with mixed zero-beta margins for estimation of taxon-taxon interactions using the normalized microbial relative abundances. Copulas allow for separate modeling of the dependence structure from the margins, marginal covariate adjustment, and uncertainty measurement. Our method shows that a two-stage maximum likelihood approach provides accurate estimation of the model parameters. A corresponding two-stage likelihood-ratio test for the dependence parameter is derived. Simulation studies show that the test is valid and more powerful than tests based upon Pearson's and rank correlations. Furthermore, we demonstrate that our method can be used to build biologically meaningful microbial networks based on the data set of the American Gut Project. \\
	
\end{abstract}

\thispagestyle{empty}
\clearpage
\pagenumbering{arabic} 

\onehalfspacing
\section{Introduction}

The past two decades have seen an increased scientific focus on understanding the structure, function, and dynamics of ``-omics" data, whether it be the genome, transcriptome, proteome, or microbiome. Specifically, the microbiome, which refers to all the microbiota and their genes in a well-defined environment, has been of interest \citep{burge_fungi_1988, lederberg_ome_2001}. Advances in technology and declining costs in sequencing gave rise to large-scale studies such as the Human Microbiome Project (HMP) and American Gut Project (AGP), which characterize the microbiome of largely healthy individuals \citep{turnbaugh_human_2007, mcdonald_american_2018}. Meanwhile, the Earth Microbiome Project (EMP) aimed to describe the uncultured diversity of the planet \citep{gilbert_earth_2014}. Much of the early research focused on microbial diversity and taxonomic classification. More recently, the focus has shifted towards differential abundance analysis and understanding how the host environment (e.g. human-host health) is associated with the microbiome \citep{mcmurdie_waste_2014, paulson_differential_2013, peng_zero-inflated_2015, scealy_regression_2011, white_statistical_2009}. From such work, it is now known that the human microbiome is associated with complex diseases, such as obesity, inflammatory bowel disease, and rheumatoid arthritis \citep{greenblum_metagenomic_2012, scher_microbiome_2011, taneja_arthritis_2014}. Whereas salinity, ecosystem type, and pH are important factors in determining soil microbial composition \citep{lozupone_global_2007, fierer_diversity_2006, thompson_communal_2017}.

Despite these advances much remains unknown about inter-microbial interactions. What is known is the micro-organisms that compose a microbiome form complex and dynamic interactions not only with their host environment, but also with one another \citep{gerber_dynamic_2014, li_microbiome_2015}. Much of the lack of information on microbial interactions comes from the fact that most standard statistical techniques in correlation or network analysis cannot be directly applied to the data, as it is sparse and compositional. For example, the commonly used Pearson's correlation is known to give spurious results for data normalized using total sum scaling \citep{pearson_mathematical_1897}.

To accommodate these limitations, novel methods to identify interactions from microbial sequencing data have been proposed. CoNet is an ensemble approach that builds a composite co-occurrence score using Pearson and Spearman correlations, as well as similarity-dissimilarity measures: Bray-Curtis and Kullback-Leibler divergence \citep{faust_microbial_2012}. Non-parametric methods, such as Maximal Information Coefficient, are used to capture linear and non-linear relationships between microbes, while binary Markov random fields are used to build interaction networks based on conditional independence \citep{reshef_detecting_2011, cai_differential_2019}. Additional methods have been developed for time series data to account for the temporal ordering of events. Lotka-Volterra models are often used to model predator-prey, or growth-decay, relationships \citep{fisher_identifying_2014, carr_use_2019}. Granger causality and extended local similarity analysis can help elucidate the directedness of such relationships \citep{ai_constructing_2019, xia_extended_2011}.

Two of the most commonly used methods, SparCC and SPIEC-EASI, build microbial networks via correlation and conditional independence metrics, respectively \citep{friedman_inferring_2012, kurtz_sparse_2015}. SparCC uses the additive log-ratio transformation and assumes that the underlying correlation of the log-ratio unobserved counts, or log-basis, it aims to estimate is sparse \citep{friedman_inferring_2012}. In contrast, SPIEC-EASI uses the centered log-ratio transformation and assumes that the network of interactions is generated from a Markov random field with a sparse inverse covariance, or precision, matrix
\citep{kurtz_sparse_2015}. While the log-ratio transformations are common in compositional data analysis, they are not particularly well suited for data with excessive zeros, as is the case with microbial sequencing data \citep{aitchison_statistical_1982}. The normality assumption of such data often does not hold and they require the use of pseudo-counts, thus forcing the assumption that the true absolute abundance for every taxa is non-zero in each sample. Furthermore, results from such transformations are difficult to interpret and are sensitive to the choice of reference group. Both SparCC and SPIEC-EASI require sparsity assumptions on the underlying dependency structure. While these assumptions may be reasonable, they are untestable. Finally, neither method provides uncertainty quantification for their estimates nor can they adjust for covariates that may influence the dependency between microbes.

As such, we propose a flexible model-based procedure to estimate the dependence between the normalized relative abundance of any two microbes. Copula models are particularly well suited for this problem as they allow for separate modeling of the univariate marginal distributions from the dependency structure. Furthermore, copulas allow for covariate adjustment in the margins and uncertainty quantification of their dependence estimate. We perform estimation on the relative abundance scale by modeling the data using a mixture of zero and beta-distribution. Such a mixture distribution has been shown to fit the microbiome relative abundance data well \citep{chen_two-part_2016,ho_metamicrobiomer_2019}. Although copula models have been widely applied to model the joint distributions with mixed margins, copula models with both marginal distributions being a mixture of discrete and continuous distributions have not been studied extensively and are the main focus of our paper. 

The remainder of this article is as follows: in the following section we review the general copula-model framework and detail a copula model with mixed zero-beta margins for microbial sequencing data. We further describe how to perform two-stage maximum likelihood estimation of all model parameters, derive their asymptotic distribution and a hypothesis test for the copula dependence parameter. We apply our method to both simulated and real data sets for comparison to Pearson's correlation, as well as rank-based Spearman's correlation and Kendall's tau. We further highlight where our method outperforms these methods.

\section{Copula Models with Mixture Margin Distributions} \label{methods}

\subsection{Zero-inflated beta marginal distribution and the copula model}
Consider a single microbial sample which can be summarized by the normalized relative abundances of the $m$-microbes, denoted by $(x_1, \dots, x_m) \in [0, 1)^m$. We assume that each $x_{j}$ follows a zero-inflated beta distribution. Accordingly, the marginal density of any given $x_{j}$ can be written as, 
\beq
f(x_j) = p_j I_j + (1 - p_j) f_{beta}(x_j ; \mu_j, \phi_j) (1 - I_j),
\eeq
where we define $p_j$ = Pr$(x_j = 0)$, $I_j$ = $I(x_j = 0)$, and 
\[
f_{beta}(x_j ; \mu_j, \phi_j) = \frac{\Gamma(\phi_j)}{\Gamma(\mu_j \phi_j) \Gamma((1 - \mu_j) \phi_j)} x_j^{\mu_j \phi_j - 1} (1 - x_j)^{(1 - \mu_j) \phi_j - 1},
\]
the density function of a beta random variable indexed by mean parameter $\mu_j$ and dispersion parameter $\phi_j$. 

It is often of interest to understand the relationship between any pair of microbes, but calculating the joint distribution of a set of non-normal random variables can be tedious and contain many parameters. As such, we propose a copula-based approach. Mathematically, a copula is the joint distribution function of a set of uniform random variables, \textbf{A} = $(A_i, \dots , A_m)$. Though, in practice, copulas can be used to describe the distribution function of any set of random variables, \textbf{X}, such that $A_k = F_k(X_k)$, where $F_k$ is the marginal cumulative distribution function of the $k^{th}$ variable, $X_k$. This is proven by Sklar's theorem, which states that any multivariate joint distribution can be described by two parts: (1) the copula function $C$ and (2) the univariate marginal distribution functions $F_k$ ($k = 1, \dots , m$) \citep{sklar_fonctions_1959}. Therefore, for any pair of microbes we can write the bivariate cumulative distribution of their normalized relative abundances as
\[
F(x_i, x_j; \boldsymbol{\gamma}_i, \boldsymbol{\gamma}_j, \theta_{ij}) = 
C(F_i(x_i; \boldsymbol{\gamma}_i), F_j(x_j; \boldsymbol{\gamma}_j); \theta_{ij}) = C(u, v; \theta_{ij}),
\]
where $U = F_i(\cdot; \boldsymbol{\gamma}_i)$ and $V = F_j(\cdot; \boldsymbol{\gamma}_j)$ are the univariate zero-inflated beta margins of $X_i$ and $X_j$, respectively, with parameters $\boldsymbol{\gamma} = (p, \mu, \phi)^\top$ and $C(\cdot; \theta)$ is a family of copula functions with dependence parameter $\theta$. The copula function links, or ties, together the margins to form the joint distribution. An advantageous property of copulas is that they completely describe the dependency between the margins via their parameter $\theta$, thus allowing for separate modeling of the margins and dependence structures. 

Moreover, we can specify a set of demographic and clinical variables that affect each microbe's presence-absence probability, mean abundance, and dispersion using a set of generalized linear models. It is for this reason that we used the alternate parameterization of the beta distribution. We assume that parameters of each margin, $p_k$, $\mu_k$, and $\phi_k$, $k  = i, j$ can be specified according to a general class of zero-inflated beta regression models as follows \citep{ospina_general_2012}:
\begin{eqnarray*}
	h_1(p_k)     =  f_1(\textbf{Q}_k, \boldsymbol{\rho}_k), 
	h_2(\mu_k) = f_2(\textbf{W}_k, \boldsymbol{\delta}_k), \mbox{and }
	h_3(\phi_k) = f_3(\textbf{Z}_k, \boldsymbol{\kappa}_k).
\end{eqnarray*}
We define $\textbf{Q}_k$, $\textbf{W}_k$, and $\textbf{Z}_k$ as the matrix of covariates of interest for the presence-absence probability, mean abundance, and dispersion of the $k^{th}$ margin, respectively; $\boldsymbol{\rho}_k$, $\boldsymbol{\delta}_k$, and $\boldsymbol{\kappa}_k$ as their corresponding vector of regression parameters; and $f_1$, $f_2$ and $f_3$ as some functions of the covariates and regression parameters. As with all GLMs, $h_1, h_2 : (0,1) \rightarrow \mathbb{R}$ and $h_3 : (0,\infty) \rightarrow \mathbb{R}$ are strictly monotonic, twice differentiable link functions. Common choices of link function for $h_1$ and $h_2$ are the logit, probit, and log-log. Likewise, the log and square-root link functions are common choices for $h_3$.

\subsection{Joint density function of bivariate copula model with two mixture marginals}
For absolutely continuous margins, the copula distribution function is unique. The joint density function of $X_i$ and $X_j$ can be found by taking mixed partial derivatives of the copula function with respect to $U$ and $V$, resulting in $f(x_i, x_j) = c(u, v; \theta_{ij})f_i(x_i)f_j(x_j)$ where $c$ is the copula density of $C$ and $f_i, f_j$ are the marginal densities of $x_i$ and $x_j$, respectively. For discrete or mixture marginals, $C$ is not unique and the calculation of the joint density function is not as straightforward.

\cite{gunawan_mixed_2020} outline a method for defining the joint density when the margins may belong to any of the three following categories: absolutely continuous, discrete, and mixtures of absolutely continuous and discrete random variables. As such, we can use this general framework to explicitly define the joint density of two zero-inflated beta random variables and the same notation for consistency. Let $\cal M$ = $\{ i, j \}$ be the index set, $\cal C (\boldsymbol{x})$ contain the indices of $\boldsymbol{x} = \{x_i, x_j\}$ with continuous $F$ at $x$, and $\cal D (\boldsymbol{x})$ = $\cal M - C(\boldsymbol{x})$ to be the set of indices of $\boldsymbol{x}$ for which $F$ has a jump point at $x$. Therefore, $\cal D$ is the null set if and only if $x_i > 0$ and $x_j > 0$. Using these two sets \cite{gunawan_mixed_2020} defines the joint density of $x_i$ and $x_j$ as:
\beq
f(x_i, x_j) = 
c_{\cal C(\boldsymbol{x})}(\boldsymbol{b}_{\cal C(\boldsymbol{x})}) 
\prod_{k \in \cal C(\boldsymbol{x})} f_k(x_k) 
\bigtriangleup_{\boldsymbol{a_{\cal D(\boldsymbol{x})}}}^{\boldsymbol{b_{\cal D(\boldsymbol{x})}}} 
C_{\cal D(\boldsymbol{x}) \vert \cal C(\boldsymbol{x})} (\cdot \vert \boldsymbol{b}_{\cal C(\boldsymbol{x})})
\eeq
Where $\boldsymbol{a} = (F_i(x_i^-), F_j(x_j^-))$ is a vector of cumulative distribution probabilities just before $x_i$ and $x_j$ and $\boldsymbol{b} = (F_i(x_i), F_j(x_j))$. Note that when $x_k > 0$, $F_k(x_k^-) = F_k(x_k)$. Moreover, $C_{\cal D(\boldsymbol{x}) \vert \cal C(\boldsymbol{x})}$ is the copula conditional distribution function of the point-masses at zero conditional on the continuous beta part and $\bigtriangleup_{\boldsymbol{a}}^{\boldsymbol{b}} g(\cdot) = \bigtriangleup_{a_i}^{b_i} \bigtriangleup_{a_j}^{b_j} g(\cdot) = g(b_i, b_j) - g(b_i, a_j) - g(a_i, b_j) + g(a_i, a_j)$. For the bivariate case this implies there are four possible scenarios:

\begin{itemize}[align=left,labelsep=1ex]
	\item \textbf{S1}: $x_i \neq 0, x_j \neq 0$, $\implies$ $\cal C$ = $\{i, j\}$, $\cal D = \emptyset$
	\begin{equation*}
	f(x_i, x_j) = c(F_i(x_i), F_j(x_j)) f_i(x_i) f_j(x_j)
	\end{equation*}
	\item \textbf{S2}: $x_i = 0, x_j \neq 0$, $\implies$ $\cal C$ = $\{j\}$, $\cal D$ = $\{i\}$ 
	\begin{equation*}
	\begin{split}
	f(x_i, x_j) &= f_j(x_j) \bigtriangleup_{F_i(x_i^-)}^{F_i(x_i)} C_{i \vert j}(\cdot \vert F_j(x_j)) \\
	& = f_j(x_j) \{ C_{i \vert j}(F_i(x_i) \vert F_j(x_j)) - C_{i \vert j}(F_i(x_i^-) \vert F_j(x_j)) \} \\
	& =  f_j(x_j) \{ C_{i \vert j}(p_i \vert F_j(x_j)) - C_{i \vert j}(0 \vert F_j(x_j)) \} 
	=  f_j(x_j) C_{i \vert j}(p_i \vert F_j(x_j))
	\end{split}
	\end{equation*}
	\item \textbf{S3}: $x_i \neq 0, x_j = 0$, $\implies$ $\cal C$ = $\{i\}$, $\cal D$ = $\{j\}$
	\begin{equation*}
	\begin{split}
	f(x_i, x_j) &= f_i(x_i) \bigtriangleup_{F_j(x_j^-)}^{F_j(x_j)} C_{j \vert i}(\cdot \vert F_i(x_i)) \\
	& = f_i(x_i) \{ C_{j \vert i}(F_j(x_j) \vert F_i(x_i)) - C_{j \vert i}(F_j(x_j^-) \vert F_i(x_i)) \} \\
	& =  f_i(x_i) \{ C_{j \vert i}(p_j \vert F_i(x_i)) - C_{j \vert i}(0 \vert F_i(x_i)) \} 
	=  f_i(x_i) C_{j \vert i}(p_j \vert F_i(x_i))
	\end{split}
	\end{equation*}
	\item \textbf{S4}: $x_i = 0, x_j = 0$, $\implies$ $\cal C = \emptyset$, $\cal D$ = $\{i, j\}$
	\begin{equation*}
	\begin{split}
	f(x_i, x_j) & = \bigtriangleup_{F_i(x_i^-)}^{F_i(x_i)} \bigtriangleup_{F_j(x_j^-)}^{F_j(x_j)} C(\cdot) \\
	&= \bigtriangleup_{F_i(x_i^-)}^{F_i(x_i)} C(\cdot, F_j(x_j)) -  C(\cdot, F_j(x_j^-))\\
	& = C(F_i(x_i), F_j(x_j)) -  C(F_i(x_i), F_j(x_j^-)) - C(F_i(x_i^-), F_j(x_j)) +  C(F_i(x_i^-), F_j(x_j^-)) \\
	& = C(p_i, p_j) -  C(p_i, 0) - C(0, p_j) +  C(0, 0) 
	= C(p_i, p_j)
	\end{split}
	\end{equation*}
\end{itemize}

The above joint distribution of $x_i$ and $x_j$ holds for any choice of copula function $C$. Although, in this paper, we chose to focus on only the Frank copula, whose properties are well suited for microbial interactions. In particular, the Frank copula can model the maximal range of dependence, meaning $\theta \in \{\ - \infty, \infty \} \setminus 0$, with $- \infty$ and $\infty$ corresponding to the Fréchet lower and upper bounds, respectively. This is particularly advantageous since other Archimedean copulas, such as the Gumbel and Joe copulas, do not permit negative dependence structures, which are likely to be seen in microbial interactions. Also, the magnitude of dependence is symmetric for positive and negative dependencies, including in the tails of the distribution. 
We use $C_{Fr}(u, v)$, $C_{v \vert u, Fr}(u, v)$ and $c_{Fr}(u,v)$ to denote the Frank copula distribution function, conditional distribution function, and joint density, respectively, where 
\beq
C_{Fr}(u, v) = -\theta \log \Bigg\{ \frac{1 +(e^{-\theta u} - 1)(e^{-\theta v} - 1) }{e^{-\theta} - 1} \Bigg\},
\eeq
and $C_{v \vert u, Fr}(u, v)$ and $c_{Fr}(u,v)$ can be derived. 
\ignore{
	\beq
	C_{v \vert u, Fr}(u, v) = \frac{e^{-\theta u} (e^{-\theta v} - 1)}{(e^{-\theta u} - 1)(e^{-\theta v} - 1) + (e^{-\theta} - 1)}
	\eeq
	
	\beq
	c_{Fr}(u,v) = \frac{-\theta (e^{-\theta} - 1)e^{-\theta (u + v)}}{ \{ (e^{-\theta u} - 1)(e^{-\theta v} - 1)  + (e^{-\theta} - 1) \}^2 }
	\eeq
}
Henceforth we assume all copulas are referring to the Frank copula.

Now that we have defined the bivariate density of $x_i$ and $x_j$ we can define the likelihood function and use a maximum likelihood estimation procedure for model parameters, $p_i, \mu_i, \phi_i, p_j, \mu_j, \phi_j,$ and $\theta_{ij}$. Using the typical full maximum likelihood estimation requires a seven-dimensional optimization procedure in the simplest case of no covariate adjustment. The numerical optimization of one function with many parameters is more difficult and computationally intensive than the numerical optimization of several functions with fewer parameters. As such, we use a two-stage, or inference-for-margins, procedure that breaks the parameter estimation into several smaller estimation problems \citep{shih_inferences_1995, joe_estimation_1996}.

\section{A Two-Stage Estimation Method and Statistical Inference}

\subsection{A two-stage estimation method}
For a sample of size $n$, with observed random vectors $\boldsymbol{X}_1, \dots, \boldsymbol{X}_n \in \R^2$ that represent the relative abundances of a pair of bacteria $(i,j)$, we consider the univariate log-likelihood functions of the zero-inflated beta margins: 
\[
\ell_k(\boldsymbol{\gamma}_k) = \sum_{l=1}^n \log f_k(x_{lk}; \boldsymbol{\gamma}_k), \quad k \in \{i,j\}
\]
and the log-likelihood function for the joint distribution,
\[
\ell(\theta, \boldsymbol{\gamma}_i, \boldsymbol{\gamma}_j) = 
\sum_{l=1}^n \log f(\textbf{X}_l; \boldsymbol{\gamma}_i, \boldsymbol{\gamma}_j, \theta).
\]

Note that we have here, and henceforth will, suppress the subscript on $\theta$, implying that we are referring to a given (\textit{i, j}) pair of microbes, unless otherwise noted. The two-stage estimation procedure \citep{shih_inferences_1995, joe_estimation_1996} can be summarized as follows:

\begin{enumerate}
	\item Assuming independence, the log-likelihoods, $\ell_i$ and $\ell_j$, of the two univariate margins are separately maximized to get estimates of their parameters $\boldsymbol{\tilde{\gamma}}_j$ and $\boldsymbol{\tilde{\gamma}}_j$, respectively.
	\item The function $\ell(\theta, \boldsymbol{\tilde{\gamma}}_i, \boldsymbol{\tilde{\gamma}}_j)$ is maximized over $\theta$ to get $\tilde{\theta}$.
\end{enumerate}

Hereafter, we denote $\boldsymbol{\eta} = (\boldsymbol{\gamma}_i, \boldsymbol{\gamma}_j, \theta)$ as the vector of all parameters, $\boldsymbol{\tilde{\eta}} = (\boldsymbol{\tilde{\gamma}}_i, \boldsymbol{\tilde{\gamma}}_j, \tilde{\theta})$ as the vector of two-stage estimators, and $\boldsymbol{\hat{\eta}} = (\boldsymbol{\hat{\gamma}}_i, \boldsymbol{\hat{\gamma}}_j, \hat{\theta})$ as the MLEs that simultaneously maximize the full log-likelihood function $\ell(\theta, \boldsymbol{\gamma}_i, \boldsymbol{\gamma}_j)$. 

We begin with the two-stage MLEs of the $k^{th}$ zero-inflated beta margin with log-likelihood $\ell_k$ equal to:
\begin{equation}
\begin{split}
	\ell_k(\boldsymbol{\gamma}_k) &= \log \Bigg\{ \prod_{l = 1}^{n} p_k^{I_{lk}} \bigg\{ (1 - p_{k}) \frac{\Gamma(\phi_k)}{\Gamma(\mu_k \phi_k) \Gamma((1 - \mu_k) \phi_k)} x_k^{\mu_k \phi_k - 1} (1 - x_k)^{(1 - \mu_k) \phi_k - 1} \bigg\}^{1 - I_{lk}} \Bigg\} \\
	&= z_k \log(p_k) + (n - z_k) \log(1 - p_k) + (n - z_k) \log \Gamma(\phi_k) - (n - z_k) \log \Gamma(\mu_k \phi_k) \\
	&\quad - (n - z_k) \log \Gamma((1 - \mu_k) \phi_k) + (\mu_k \phi_k - 1) \sum_{l=1}^{n} \log(x_{lk}) + ((1 - \mu_k) \phi_k - 1) \sum_{l=1}^{n} \log(1 - x_{lk}) 
\end{split} \label{loglik-k}
\end{equation}
Where $z_k = \sum_{l=1}^{n} I_{lk} = \sum_{l=1}^{n} I(x_{lk} = 0)$ is the number of observations with $x_k = 0$ and $\Gamma$($W+1$) = $W!$ is the gamma function. 
\ignore{We take the first derivative of \ref{loglik-k} with respect to $\boldsymbol{\rho}_k, \boldsymbol{\delta}_k,$ and $\boldsymbol{\kappa}_k$ to get the following score equations: 
	\begin{eqnarray}
	\boldsymbol{U}_{\boldsymbol{\rho}_k}     &= \frac{\partial \ell_k(\boldsymbol{\gamma}_k)}{\partial p_k} \frac{\partial p_k}{\partial \boldsymbol{\rho}_k} \\
	\boldsymbol{U}_{\boldsymbol{\delta}_k}   &= \frac{\partial \ell_k(\boldsymbol{\gamma}_k)}{\partial \mu_k} \frac{\partial \mu_k}{\boldsymbol{\delta}_k} \\ 
	\boldsymbol{U}_{\boldsymbol{\kappa}_k} &= \frac{\partial \ell_k(\boldsymbol{\gamma}_k)}{\partial \phi_k} \frac{\partial \phi_k}{\boldsymbol{\kappa}_k} 
	\end{eqnarray} \label{Uk}
	Setting the score equations equal to zero do not give a closed from solution. Instead}
We use the Newton-Raphson algorithm to numerically find the MLEs of $\boldsymbol{\rho}_k, \boldsymbol{\delta}_k,$ and $\boldsymbol{\kappa}_k$.

Now that the marginal two-stage MLEs, $\boldsymbol{\tilde\gamma}_{k}$, have been defined they can be plugged into the full likelihood $\ell(\theta, \boldsymbol{\gamma}_i, \boldsymbol{\gamma}_j)$ to give:
\begin{equation}
\begin{split}
	\ell(\theta, \boldsymbol{\tilde{\gamma}}_i, \boldsymbol{\tilde{\gamma}}_j) &\propto 
	\sum_{l \in S1} \log \{ -\theta (e^{-\theta} - 1) \} + \sum_{l \in S1} \log \{ e^{-\theta(\tilde{u}_l + \tilde{v}_l)} \} \\ 
	&- \sum_{l \in S1} 2 \log \{ (e^{-\theta \tilde{u}_l} - 1) (e^{-\theta \tilde{v}_l} - 1) + (e^{-\theta} - 1) \} \\
	&+ \sum_{l \in S2} \log \Bigg\{ \frac{e^{-\theta \tilde{p_i}} - 1}{(e^{-\theta \tilde{p_i}} - 1) (e^{-\theta \tilde{v}_l} - 1) + (e^{-\theta} - 1)} \Bigg\} 
	+ \sum_{l \in S2} \log \{ e^{-\theta \tilde{v}_l} \} \\
	&+ \sum_{l \in S3} \log \Bigg\{ \frac{e^{-\theta \tilde{p_j}} - 1}{(e^{-\theta \tilde{u}_l} - 1) (e^{-\theta \tilde{p_j}} - 1) + (e^{-\theta} - 1)} \Bigg\} 
	+ \sum_{l \in S3} \log \{ e^{-\theta \tilde{u}_l} \} \\
	&+ \sum_{l \in S4} \log \Bigg\{ -\theta \log \Bigg\{ 1 + \frac{(e^{-\theta \tilde{p_i}} - 1) (e^{-\theta \tilde{p_j}} - 1)}{e^{-\theta} - 1} \Bigg\} \Bigg\}
\end{split} \label{ts-loglik}
\end{equation}
The log-likelihood can be split into four parts, each corresponding to the contribution of observations from one of the four scenarios given previously. The notation $\sum_{l \in S1}$ implies summation over all the observations that fall into the first scenario, $x_i \neq 0$ and $x_j \neq 0$, likewise for other summations. Moreover, $\tilde{u_l} = F_i(x_{li}; \boldsymbol{\tilde\gamma}_i)$ is the cumulative distribution function of microbe $i$ evaluated at $x_{li}$ with the two-stage MLEs plugged in for the marginal parameters. The same holds for $\tilde{v_l}$ and microbe $j$.
The two-stage MLE of $\theta$ is found  numerically using Brent's method \citep{brent_algorithms_2013}.

\subsection{Asymptotic normality}

\cite{joe_asymptotic_2005} obtained the asymptotic covariance matrix for the two-stage estimator $\boldsymbol{\tilde{\eta}}$ using the theory of inference functions. Specifically, by defining the inference functions
\beq
\boldsymbol{g} = (g_i, g_j, g_{\theta})^\top,
\eeq
where
\beq
g_k = \frac{\partial \log f_k(\cdot; \boldsymbol{\gamma}_k)}{\partial \boldsymbol{\gamma}_k}, \quad \text{for } k \in \{i,j\}
\eeq
and $g_{\theta}=\partial \log f(\cdot;\boldsymbol{\eta})/\partial \theta$, it is shown that
\beq
\sqrt{n} (\boldsymbol{\tilde{\eta}} - \boldsymbol{\eta}) \xrightarrow{D} MVN(0,\textbf{V}), \quad \text{as $n\to \infty$,}
\eeq
where  $\boldsymbol{V} = (-\boldsymbol{D}_g^{-1}) \boldsymbol{M}_g (-\boldsymbol{D}_g^{-1})^\top$, 
$\boldsymbol{M}_g = \text{Cov}(\boldsymbol{g}(Y \vert \boldsymbol{\eta})) = \E [\boldsymbol{gg}^\top]$, 
and $\boldsymbol{D}_g = \E[\partial \boldsymbol{g}(Y, \boldsymbol{\eta})/\partial \boldsymbol{\eta}^\top]$.

Now let $\mathcal{J} = \text{Cov}(\boldsymbol{g}_i, \boldsymbol{g}_j) = \E [\boldsymbol{g}_i \boldsymbol{g}_j^\top]$,  $\mathcal{I}=-\E [\partial^2 \log f/\partial \boldsymbol{\gamma}_i \partial \boldsymbol{\gamma}_j^\top]$ and $\mathcal{I}_{k\theta}=-\E [\partial^2 \log f/\partial \boldsymbol{\gamma}_k \partial \theta]$ for $k=i,j$. Then

\beq
-D_g =
\begin{bmatrix}
	\mathcal{J}_{ii} & 0                      & 0\\
	0                      & \mathcal{J}_{jj} & 0\\
	\mathcal{I}_{\theta i} & \mathcal{I}_{\theta j} &\mathcal{I}_{\theta\theta}
\end{bmatrix}
,\quad -D_g^{-1} = 
\begin{bmatrix}
	\mathcal{J}^{-1}_{ii} & 0                              &0 \\
	0                              & \mathcal{J}^{-1}_{jj} & 0 \\
	a_i                            & a_{j}                        & \mathcal{I}^{-1}_{\theta\theta}
\end{bmatrix}
,\quad M_g = 
\begin{bmatrix}
	\mathcal{J}_{ii} &\mathcal{J}_{ij}  &0\\
	\mathcal{J}_{ji} & \mathcal{J}_{jj} & 0\\
	0                      & 0                     & \mathcal{I}_{\theta\theta}
\end{bmatrix}.
\eeq
where $a_k= -\mathcal{I}_{\theta\theta}^{-1}\mathcal{I}_{\theta k} \mathcal{J}_{kk}^{-1}$ for $k=i,j$. 

\subsection{A re-scaled likelihood ratio test}

In general, we are interested in determining if any two microbes $i$ and $j$ have a dependence structure such that $\theta = \Theta_{0}$ for some pre-specified $\Theta_{0}$. We propose a re-scaled likelihood ratio test to do so. Consider the general hypothesis testing problem: 
\[
H_0: \theta \in \Theta_0, \quad \text{vs.} \quad H_1: \theta \in \Theta_1
\] 

Suppose $\ell = (\ell_i, \ell_j, \ell_{\theta})^\top$ where $\ell_k = \log f_k(\cdot; \boldsymbol{\gamma}_k)$ for $k= i,j$, and $\ell_{\theta} = \log f(\cdot; \boldsymbol{\eta})$. Define the two-stage likelihood ratio test statistic as:
\beq
\Lambda' = -2 \omega [\ell(\theta_0, \boldsymbol{\tilde{\gamma}}_i, \boldsymbol{\tilde{\gamma}}_j) - \ell(\tilde{\theta}, \boldsymbol{\tilde{\gamma}}_i, \boldsymbol{\tilde{\gamma}}_j)],
\eeq
where
\[
\omega = \bigg( 1 + \mathcal{I}_{\theta\theta}^{-1}
(\mathcal{I}_{\theta 1} \mathcal{J}_{11}^{-1} \mathcal{I}_{1 \theta} + 
\mathcal{I}_{\theta 2} \mathcal{J}_{22}^{-1} \mathcal{I}_{2 \theta} + 
\mathcal{I}_{\theta 1} \mathcal{J}_{11}^{-1} \mathcal{J}_{12} \mathcal{J}_{22}^{-1} \mathcal{I}_{2 \theta} + 
\mathcal{I}_{\theta 2} \mathcal{J}_{22}^{-1} \mathcal{J}_{21} \mathcal{J}_{11}^{-1} \mathcal{I}_{1 \theta} )  \bigg)^{-1}.
\]
%

\bet
Under standard regularity conditions, we have $\Lambda' \xrightarrow{D} \chi_1^2$.
\eet

It can be shown that the above two-stage likelihood ratio test is equivalent to the pseudo-likelihood ratio test \citep{liang_asymptotic_1996}.

Most often, the hypothesis we are interesting in testing is $\Theta_{0} = \theta_{I}$ where $\theta_I$ is the value of the dependence parameter that corresponds to the independence copula. For the Frank copula this is equivalent setting $\theta_{I} = 0$. Under independence, it can be shown that $\mathcal{I}_{1 \theta} = \mathcal{I}_{2 \theta} = 0$, implying that $\tilde{\theta}$ is asymptotically efficient and the two-stage likelihood ratio statistic reduces to the regular LRT statistic \citep{shih_inferences_1995, genest_semiparametric_1995}.

\section{Simulation Studies}

Simulation studies were used to assess the bias and variance of the two-stage estimation procedure, as well as the Type I error and power of the two-stage likelihood ratio test. The data was simulated using the Rosenblatt transformation, a variant of the probability integral transformation. Let $U$ and $V$ be defined as earlier and define a new random variable $W$ such that, 
\[ W = C_{v \vert u} (u, v) \coloneqq \frac{\partial C(u, v)}{\partial u} = Pr(V = v \vert U = u). \]

By the Rosenblatt transformation, $U$ and $W$ are independent uniform random variables and we can define the following simulation algorithm for any two microbes: 

\begin{enumerate}
	\item Simulate $U \sim$ Uniform(0,1) and $W \sim$ Uniform(0,1)
	\item Solve for $v$ using: \\ 
	$w = C_{v \vert u}(u,v) = \frac{e^{- \theta u} (e^{- \theta v} -  1)}{(e^{- \theta} - 1) + (e^{- \theta u}- 1)(e^{- \theta v} - 1) }
	\Leftrightarrow
	v = - \frac{1}{\theta} \log \big\{ 1 + \frac{w(e^{- \theta} - 1)}{w + e^{- \theta u}(1 - w)} \big\}$
	\item Solve for $x_i$ using the definition of $U$: \\
	$u = F_i(x_i) \Leftrightarrow x_i = F_i^{-1}(u) = 
	\begin{cases}
	0 \quad &\text{if } u \leq p_i \\
	F_{beta}^{-1} \big(\frac{u - p_i}{1 - p_i} \big) \quad &\text{if } u > p_i
	\end{cases}$ \\
	Likewise, the procedure for $x_j$ and $V$ is the same.
\end{enumerate}

The process above is repeated for a sample size of $n$. In the event that the simulation scheme above results in less than three non-zero relative abundances for either microbial taxa the procedure is repeated. This is because at least three non-zero observations are needed to be able to estimate the three taxa-specific marginal parameters. Additionally, for any simulated data set, if the two taxa are mutually exclusive, meaning no pair of observations have non-zero relative abundance for both taxa, or if only one pair of observations has non-zero relative abundance for both taxa, the procedure is repeated. This was done because such scenarios lead to dependence parameters hitting the lower boundary of estimation and/or cause unstable variance estimates. 

Simulations are performed under a variety of marginal parameter settings to understand the robustness of the estimation procedure. The dependence parameter $\theta$ was selected from \{-2.5, -1, 0, 0.5, 1.5, 3\}. Under the marginal settings of no covariate adjustment the zero-inflation probabilities, $(p_i, p_j)$, were selected from \{(0.10, 0.25), (0.40, 0.50), (0.60, 0.75), (0.20, 0.75)\} and the parameters of the beta portion of the marginal distributions, $(\mu_k, \phi_k),\ k = i,j$, were selected from $\big\{ \big( \frac{2}{7}, 7 \big), \big( \frac{5}{7}, 7 \big), \big( \frac{1}{2}, 4 \big), \big( \frac{1}{3}, 9 \big), \big( \frac{2}{3}, 9 \big), \big( \frac{1}{2}, 6 \big) \big\}$. 

We also performed simulation with a single continuous covariate affecting the presence-absence probability of each microbe. Under this setting we assumed that both $Q_{i1}$ and $Q_{j1}$ are drawn from a standard normal distribution. With corresponding vectors of true regression coefficients $ \{ \boldsymbol{\rho}_i, \boldsymbol{\rho}_j \}$ assumed to be from one of the three following settings: $\{ (-0.5, 0.7)^\top$, $(-0.3, 0.4)^\top \}$, $\{ (-0.1, 0.7)^\top$, $(0.1, 0.4)^\top \}$, and $\{ (0.5, 0.7)^\top$, $(0.8, 0.4)^\top \}$. In general these models correspond to low-low, low-high, and high-high zero-inflation probabilities, respectively. The mean abundances are specified as $\mu_i = \frac{e^{-0.7}}{1 + e^{-0.7}}$ and $\mu_j = \frac{e^{-1}}{1 + e^{-1}}$ and the dispersion parameters as $\phi_i = \phi_j = e^{1.5}$. For each of the parameters settings combinations the sample size was set to $n = 50$. Under the setting with no covariates and independence (i.e. $\theta$ = 0) additional simulations were run for a larger sample size of 250. All simulations were repeated $m = 500$ times.

\subsection{Parameter estimation}
\begin{figure}[!ht]
	\begin{center}
		\includegraphics[width=15cm,height=0.49\textheight]{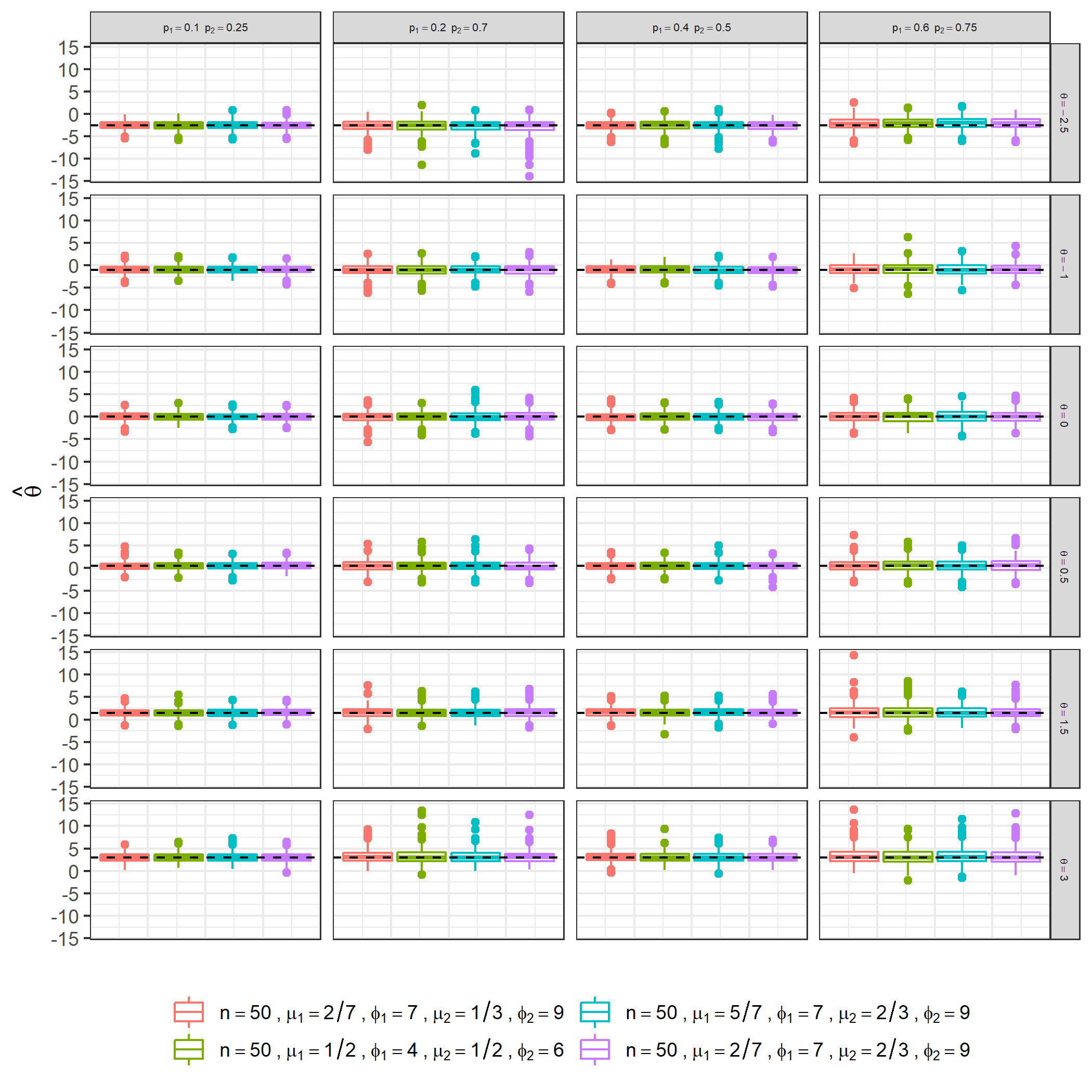} \\
		\includegraphics[width=15cm,height=0.40\textheight]{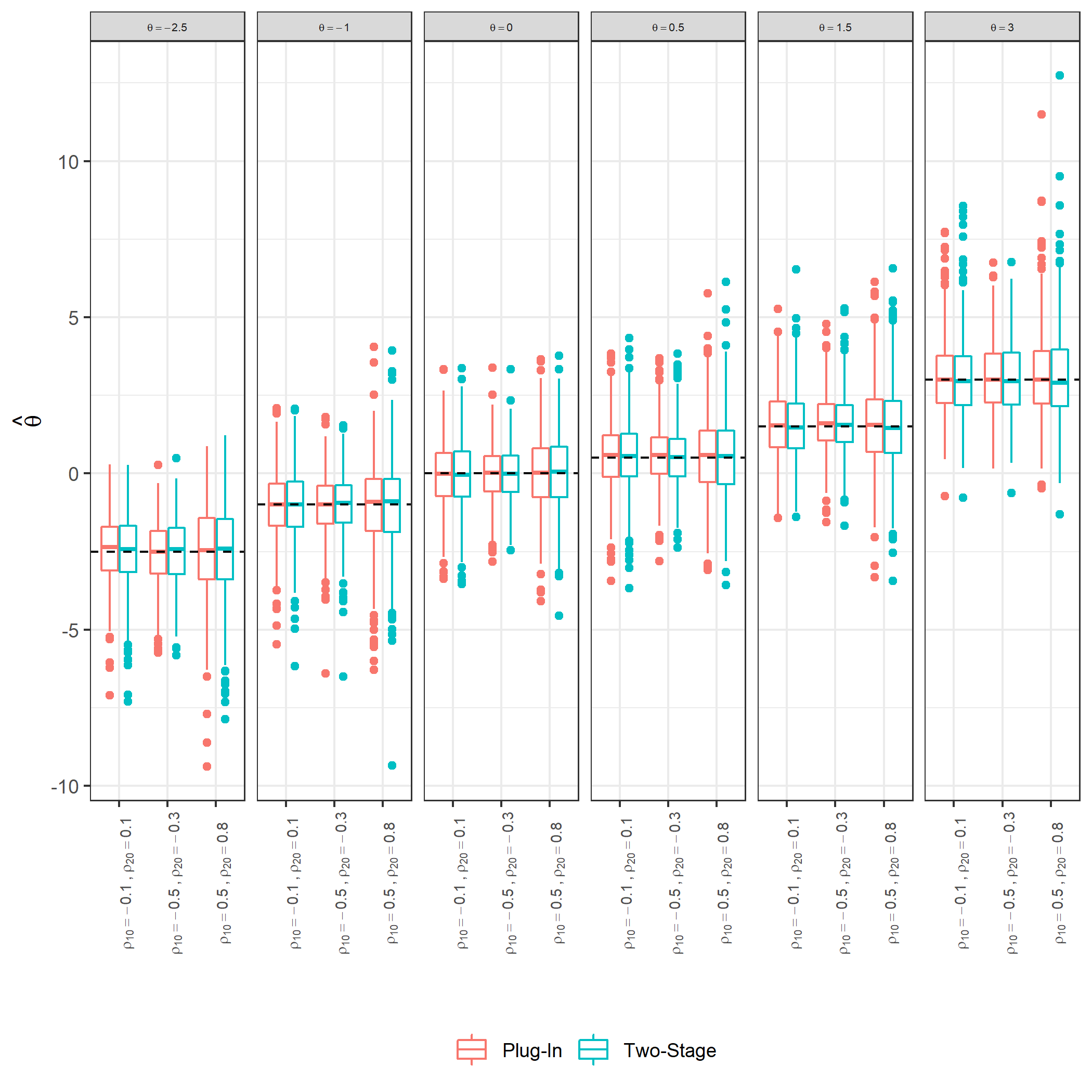}
	\end{center}
	\caption{Boxplot of estimated $\tilde\theta$ values across 500 simulations. The black dashed line represents the true $\theta$ value. Data was simulated under varying strength of dependence, mean, dispersion and zero-inflation parameter settings, without (top panel) and with (bottom panel) covariate adjustment.}
	\label{fig:consistencyNC50}
\end{figure}

The two-stage estimator is unbiased under all dependence, zero-inflation, and marginal parameter settings (Figure \ref{fig:consistencyNC50}). However,   under high zero-inflation, we observe some larger outliers in the estimates. This is expected since too many zeros in the data can lead to an unstable estimate of the parameters.  

In addition to estimating $\theta$ we also calculate its variance. The number of the second derivatives necessary to calculate the covariance matrix, \textbf{V}, is large making it analytically difficult to do so. Therefore, we replace it with a consistent estimator, such as the jackknife estimator:
\[
n^{-1} \tilde{V} = \sum_{l = 1}^{n} \big( \boldsymbol{\tilde{\eta}}^{(l)} - \boldsymbol{\tilde{\eta}})^\top (\boldsymbol{\tilde{\eta}}^{(l)} - \boldsymbol{\tilde{\eta}} \big).
\]

\begin{figure}[!ht]
	\begin{center}
		\includegraphics[width=15cm,height=0.48\textheight]{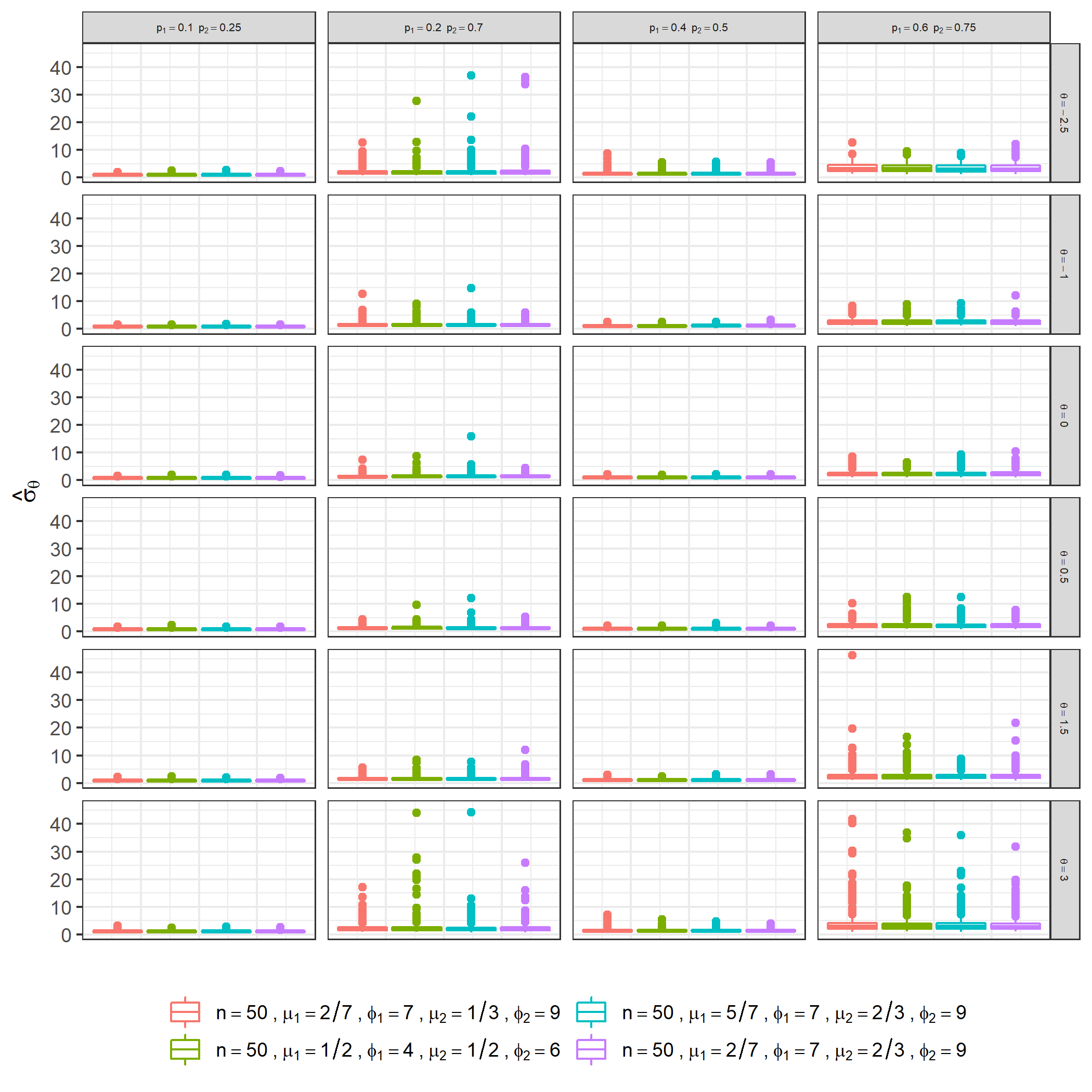} \\
		\includegraphics[width=15cm,height=0.40\textheight]{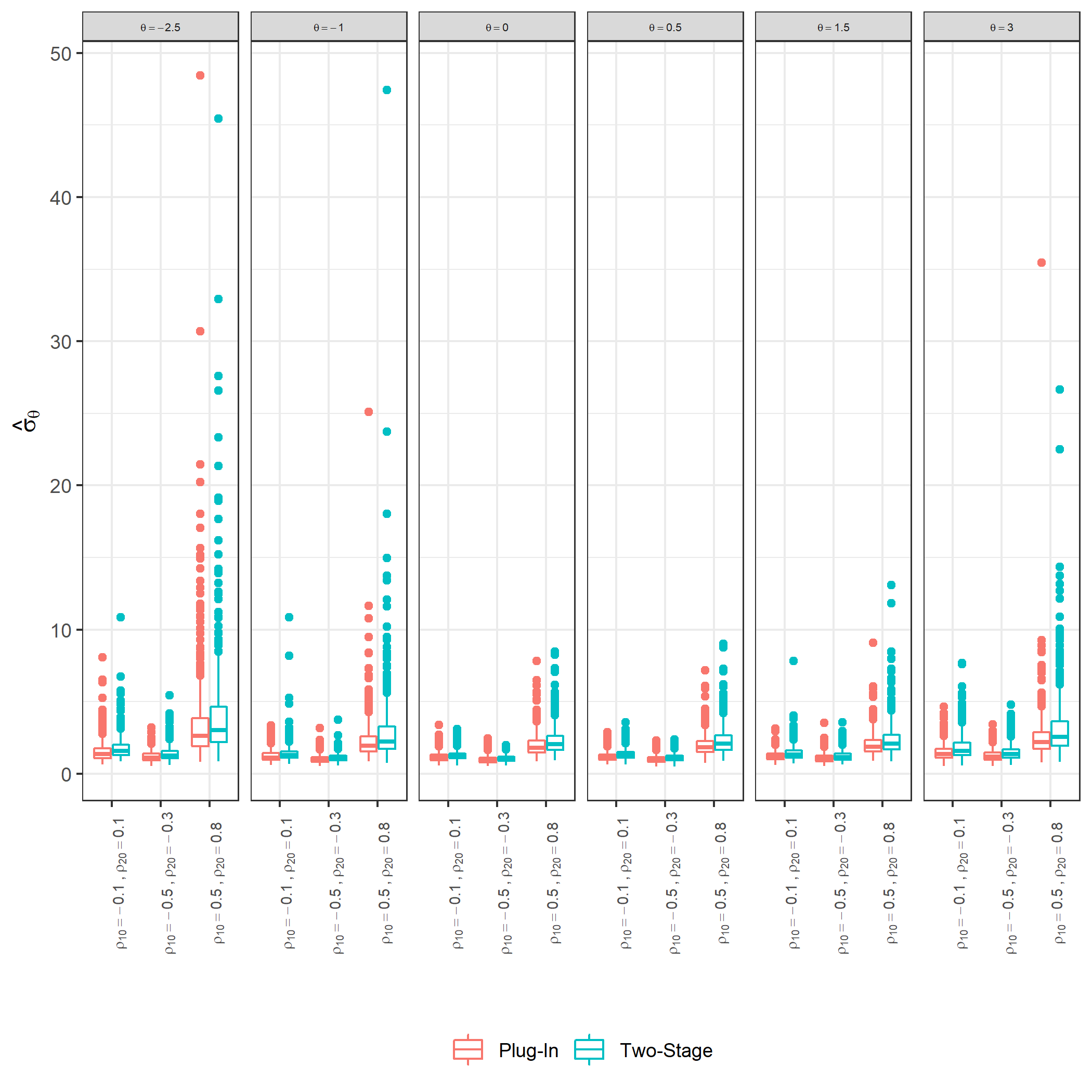}
	\end{center}
	\caption{Boxplot of estimated jackknife variance of $\tilde\theta$, denoted as $\hat\sigma_{\theta}^{2}$, across 500 simulations. Data was simulated under varying strength of dependence, mean, dispersion and zero-inflation parameter settings, without (top panel)  and with (bottom panel) covariate adjustment. Outliers with variance values greater than 50 were removed from plots.}  
	\label{fig:varNC50}
\end{figure}

\begin{figure}[!ht]
	\begin{center}
		\includegraphics[width=15cm,height=0.60\textheight]{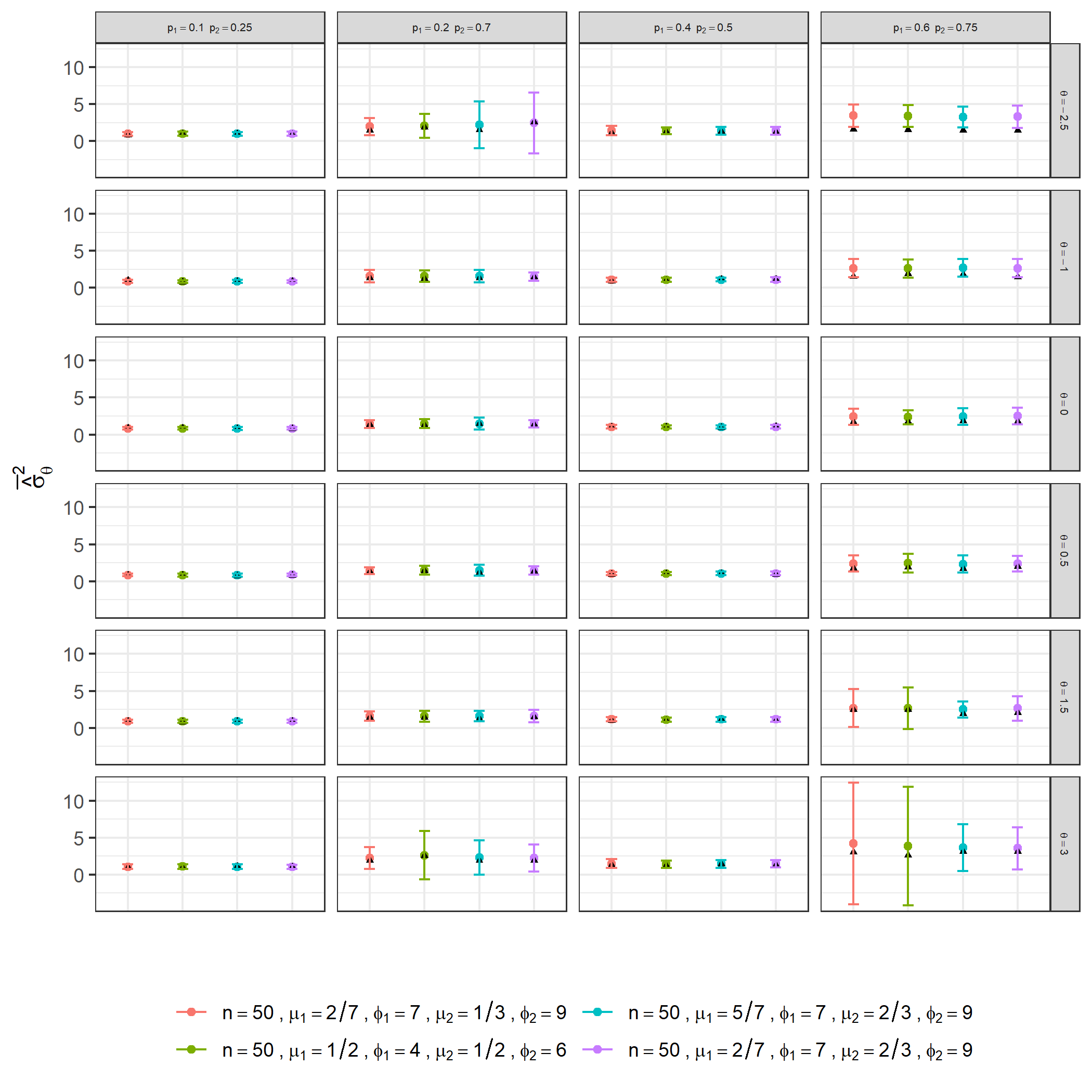}
	\end{center}
	\caption{Mean and standard error (bars) of the estimated jackknife variance of $\tilde\theta$ from data simulated under varying strength of dependence, mean, dispersion and zero-inflation parameter settings, without covariate adjustment. Black triangles correspond the empirical (sample) variance.}
	\label{fig:varBiasNC50}
\end{figure}

The variance of $\theta$ is the $(7,7)^{th}$ entry of $n^{-1}\tilde{\textbf{V}}$, denoted as $\hat\sigma_\theta^2$, and $\boldsymbol{\tilde{\eta}}^{(l)}$ is a vector of two-stage maximum-likelihood estimates calculated with the $l^{th}$ observation removed. In general, the variance increases as zero-inflation increases, regardless of dependence or marginal parameter values (Figure \ref{fig:varNC50}). Specifically, without adjusting for covariates, under high zero-inflation of both microbes and moderate-to-strong positive dependence, there is an increase in large outlier estimates. These results  show that the mean of the analytical variance is typically larger than the empirical (sample) variance of $\tilde\theta$ across all 500 simulations (Figure \ref{fig:varBiasNC50}). Though the latter almost always falls within the standard error of the former. The difference between the two increases with zero-inflation. This indicates that the jackknife estimator is conservative (upwardly biased) and may lead to a two-stage likelihood ratio test that is conservative as well. As to be expected, as the sample size increases the variance decreases across the board, though the same trends are seen (results not shown). 

\subsection{Type I error and power}

\begin{figure}[!ht]
	\begin{center}
		\includegraphics[width=15cm,height=0.5\textheight]{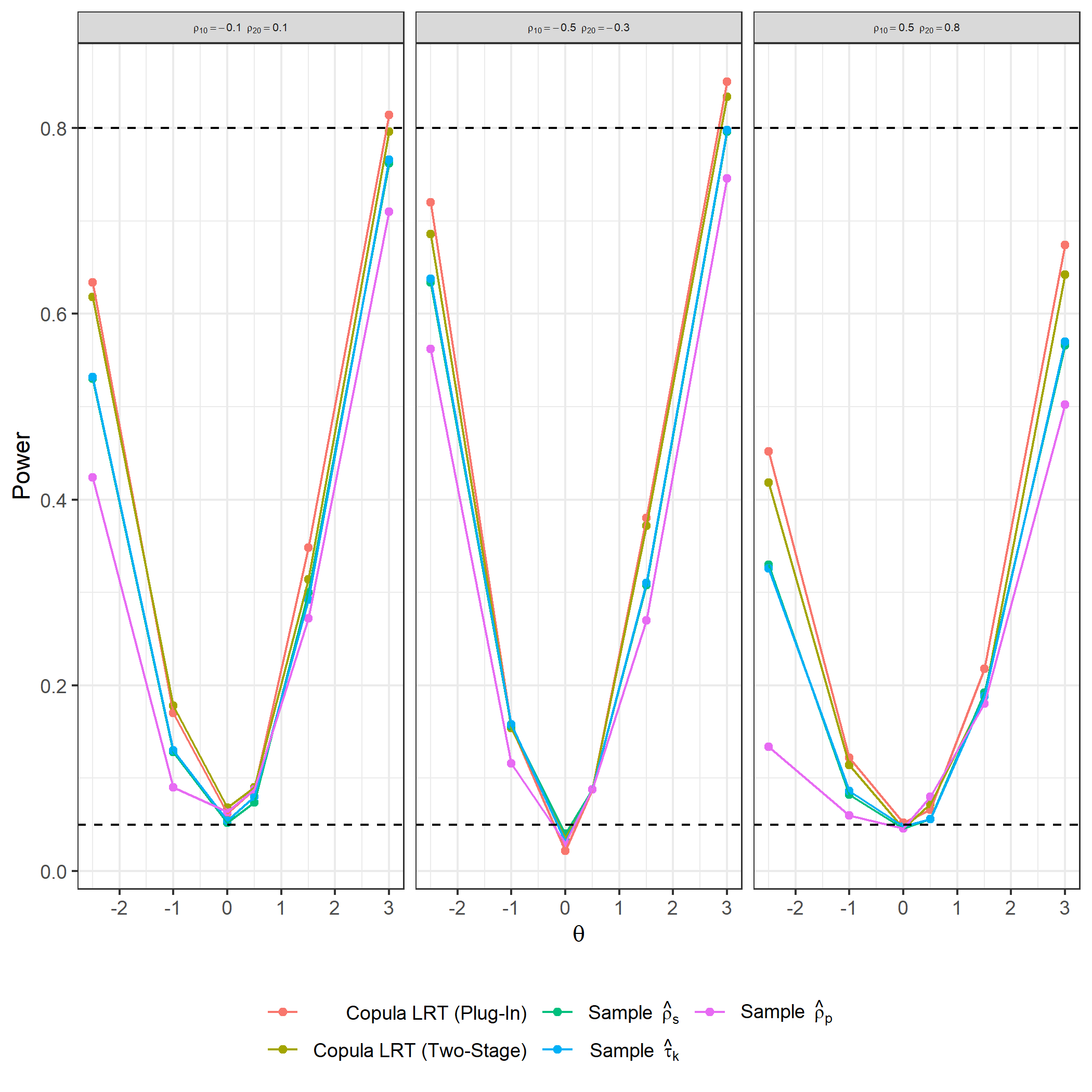}
	\end{center}
	\caption{Power curves of the likelihood ratio test for independence using the Frank copula model with zero-inflated beta margins and a two-stage maximum likelihood estimation procedure. Black dashed lines represent 80\% power and 5\% Type I error. Power was calculated under $n=50$, varying strength of dependence, and zero-inflation probabilities affected by one continuous covariate.}
	\label{fig:power1Q}
\end{figure}

We are also interested in assessing the Type I error and power of the two-stage likelihood ratio test. Specifically, we would like to test the null hypothesis that microbes $i$ and $j$ are independent (\textit{i.e.} $H_0: \theta = 0$ for the Frank copula) versus the general two-sided alternative hypothesis that microbes $i$ and $j$ are not independent (\textit{i.e.} $H_1: \theta \neq 0$). For the setting where one continuous covariate is influencing the zero-inflation probability, our proposed likelihood ratio test for independence uniformly outperforms sample correlation tests for independence using Pearson's, Spearman's and Kendall's tau rank correlation (Figure \ref{fig:power1Q}). Under low to moderate zero-inflation, as the absolute value of true $\theta$ moves away from zero, in either direction, the power of the test increases symmetrically. This does not hold under dual-high zero-inflation where the power to detect a true positive dependence structure increases much more rapidly than that of a true negative dependence structure. This trend does not hold in the setting without covariates (results not shown), under which the four tests perform comparably. This is likely due to the unique mapping between $\theta$ and Spearman's and Kendall's tau rank-based correlations in such settings. Though, there is a slight improvement in our proposed method under dual-high zero inflation. Correspondingly, under this particular setting the sample-based correlation estimates are biased towards the zero, compared to their unbiased copula-based estimates.

\section{Analysis of Microbial Network in Healthy Human Gut}

\subsection{Pairwise microbial dependence estimation}

We applied our method to data from the American Gut Project (AGP), a self-selected, open-platform cohort \citep{mcdonald_american_2018}. The cohort consists of individuals mostly from the United States, with some from the United Kingdom and Australia, who opted into the study by providing informed consent and paying a fee to offset the cost of processing and sequencing. The data, both 16S rRNA gene sequencing and self-reported meta-data, are publicly available in The European Bioinformatics Institute repository under the accession ERP012803.

The data consisted of fecal microbiome samples from 3679 citizen-scientists and 971 unique genera. We filtered the sequencing data such that any reads that were unassigned at the genera level were removed. Any genera with a prevalence of less than 20\% across all subjects were removed as well. This left a total of 68 genera for downstream analyses. Furthermore, any samples that had total number of reads of zero after the aforementioned filtering were removed. Since the data also included self-reported meta-data we choose to adjust for covariates known to influence the composition of the gut microbiome in the marginal zero-inflated beta regression models. In particular, we adjusted for age (44.6 years $\pm$ 17.4), bmi (23.9 $\pm$ 5.26) and antibiotic use (69\% not in the last year, 14\% in the last year, 13\% in the last six months, 2\% in the last month, and 2\% in the last week). Due to the low rate of missing data for each, $<$ 5\% for age and antibiotic use and about 10\% for BMI, we performed a complete case analysis. We further restricted our sample of interest to ``healthy" individuals, defined as those who reported not having inflammatory bowel disease or diabetes, as both are known to be associated with dysbiosis. This left 2754 samples remaining.

From these 68 genera we can form 2278 unique pairs. For each of these pairs, we perform two-stage maximum likelihood estimation of the parameters and a likelihood ratio test for independence. Due to the large number of pairwise tests, we adjust for multiple comparisons by controlling the false discovery rate at 1\% level. In particular, since the test statistics are not independent from one another we use the Benjamini-Yekutieli procedure \citep{benjamini_control_2001}. After FDR control we identify 1314 pairs of taxa with a significant dependence among healthy subjects.

\subsection{Properties of microbial network in healthy human gut}

\begin{figure}[!ht]
	\centering
	\includegraphics[width=15cm]{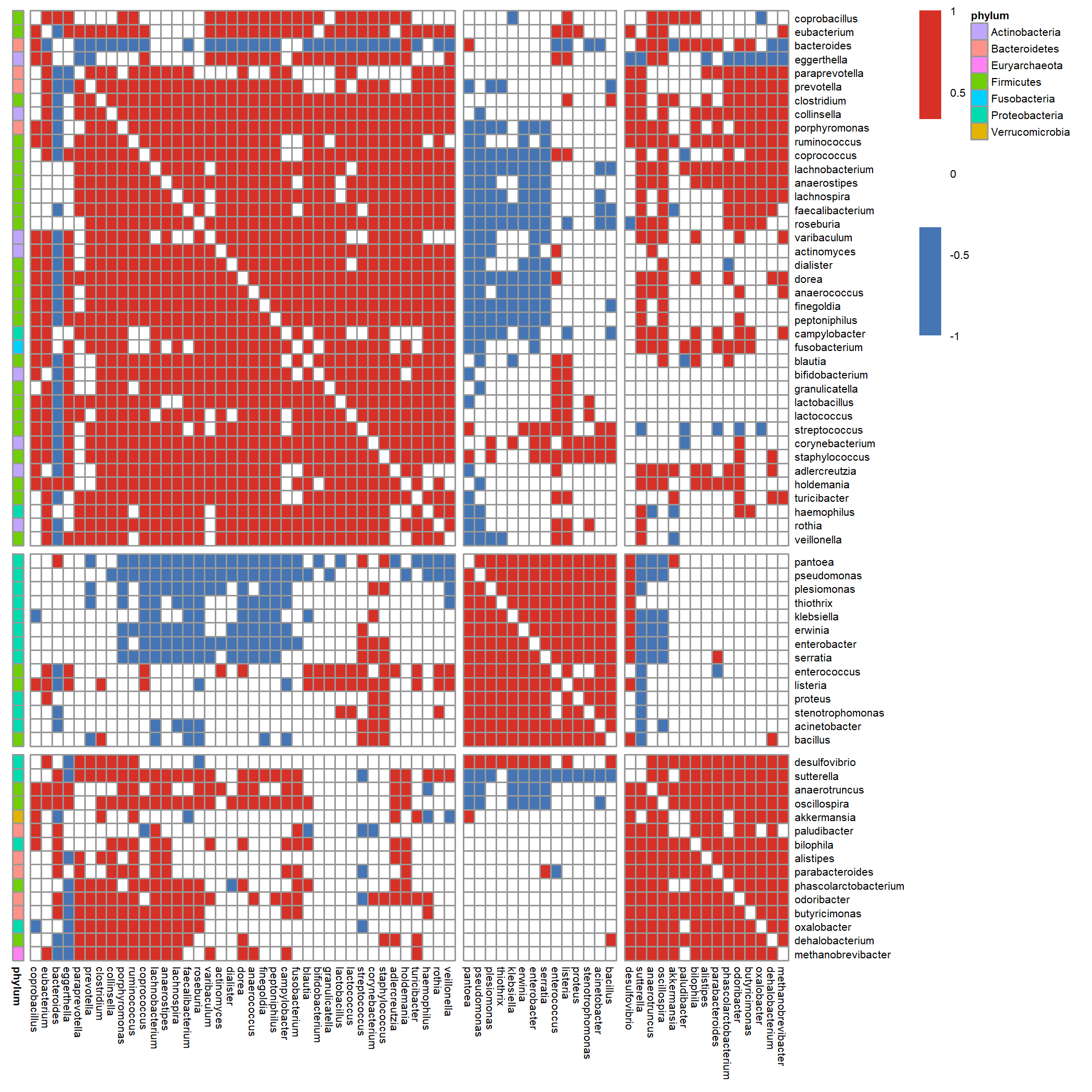}
	\caption{Heatmap of the AGP adjacency matrix with the dendrogram from complete agglomerative hierarchical clustering.}
	\label{fig:heatmapAdj}
\end{figure}

We use the results from the likelihood ratio test for independence to construct an adjacency matrix and perform network analysis. More specifically, two microbes are said to have a connection if the result from their test for independence was significant (FDR-controlled p-value is $< 0.01$), otherwise two microbes are said to be unconnected. A heatmap of the complete agglomerative hierarchical clustered adjacency matrix shows the relationship between the microbial pairs (Figure \ref{fig:heatmapAdj}). Moreover, the adjacency matrix can be represented in network form with each microbial genera as a node and each significant pair as an edge. Figure \ref{fig:heatmapAdj} shows that the network consists mostly of pairs with positive dependence, especially within clusters, with some negative dependencies between a small set of taxa, mostly between clusters. Furthermore, the nodes of the network form three distinct clusters, identified by a cutting the hierarchical clustering dendrogram. The most common phylum in each cluster was Firmicutes, Proteobacteria, and Bacteroidetes , respectively. This implies that the clusters have a biological interpretation with taxa of the same phylum tending to be members of the same cluster.

To summarize the resulting network, we  calculate  the average of some network summary statistics measures, including average degree of 0.577 (sd=0.146), average closeness of 0.710 (sd=0.072), average betweenness of 0.006 (sd=0.004).  The high average degree of the nodes implies the network is dense with many connections. This is further implied by the network's edge density of 0.58. Meanwhile the high eigenvalue centrality of 0.704 (sd=0.198) implies that well connected nodes are likely to be connected with each other. The network also has a diameter of 2 and a mean distance of 1.42. 

We simulated 1000 random graphs from the Erd\H{o}s--R\'{e}nyi model with the same number of links as the AGP network and compared global network measures from these graphs to that of the AGP network.  Both the average cluster coefficient (0.695) and modularity (0.137) of the AGP network were significantly different from those of the random graphs ($p<0.001$). Thus implying that the network structure and clusters are not formed due to random noise in the data. Additionally, we compared the cumulative degree distribution of the AGP network to that of the 1000 random graphs ($p=0.077$). We observe that  distribution of the random graphs are begins around 35 degrees and increases steeply until it levels off at 50 degree. In contrast, the distribution of the  AGP network begins early around 20 degrees and rises slowly until a maximum of approximately 60 degrees.

\subsection{Stability analysis}
To assess the robustness of the identified microbial pairs to slight changes in the observed data we took 50 bootstrap samples of the relative abundance data, then repeated the estimation and testing analyses. If the identified bivariate pairs are truly associated with one another we should see high stability, or overlap, in the identified pairs between the original data and bootstrap samples. The average number of significant dependent pairs of taxa, after FDR control using the BY procedure at the 0.01 significance level, across all bootstrap samples rounded to the nearest integer is 1335. The minimum number of identified pairs is 1274 and the maximum is 1393. The average overlap and dice coefficients between the pairs identified in the original data and those of each bootstrap sample is 0.940 (sd=0.010) and 0.930 (sd=0.006), respectively. Thus indicating that the identified significant pairs are robust to small changes in the observed data. Furthermore, of the 1314 microbial pairs identified from original data, 875 of these pairs were also identified in all 50 bootstrap samples and 1071 pairs were identified in over 90\% of them. Only 14 were identified in less than half of the bootstrap samples.

\section{Discussion}

In this paper we described a bivariate copula-based density for microbial relative abundance data using zero-inflated beta margins. As such, this allowed for a two-stage maximum likelihood estimation and corresponding two-stage likelihood ratio test for the copula dependence parameter. Performing estimation and inference on the relative abundance scale avoids strict sparsity assumptions necessary when using the unobserved absolute abundances \citep{friedman_inferring_2012, kurtz_sparse_2015}. While using model based method allows for covariate adjustment via the margins and uncertainty quantification of the dependence parameter.

The low bias and high efficiency of the proposed two-stage estimator of the dependence parameter under unknown margins is a valid, and less computationally intensive, alternative to full maximum likelihood estimation. We extend current work on copula models with mixed margins \citep{gunawan_mixed_2020}, as well as work on copula two-stage estimation \citep{shih_inferences_1995, joe_asymptotic_2005} with our proposed two-stage likelihood ratio test. Simulation studies show under the independence hypothesis the test controls Type I error and is more powerful than tests based on sample correlation measures. 

While this paper focuses on the Frank copula, the methods are quite general and hold for any Archimedean copula. Extensions of this work include goodness-of-fit test to compare copula choice. For example, both the $t$- and Clayton copula can model positive and negative dependence, but they assume tail dependence which Frank does not. Additional extensions include modifications to handle longitudinal data in order to understand the changes of microbial dynamics.

\singlespacing
\bibliographystyle{apa}
\bibliography{mmc-bibliography}

\begin{thebibliography}{}

\bibitem[\protect\astroncite{Ai et~al.}{2019}]{ai_constructing_2019}
Ai, D., Li, X., Liu, G., Liang, X., and Xia, L.~C. (2019).
\newblock Constructing the {Microbial} {Association} {Network} from large-scale
  time series data using {Granger} causality.
\newblock {\em Genes}, 10(3):216.

\bibitem[\protect\astroncite{Aitchison}{1982}]{aitchison_statistical_1982}
Aitchison, J. (1982).
\newblock The {Statistical} {Analysis} of {Compositional} {Data}.
\newblock {\em Journal of the Royal Statistical Society: Series B
  (Methodological)}, 44(2):139--160.

\bibitem[\protect\astroncite{Benjamini and
  Yekutieli}{2001}]{benjamini_control_2001}
Benjamini, Y. and Yekutieli, D. (2001).
\newblock The control of the false discovery rate in multiple testing under
  dependency.
\newblock {\em Annals of Statistics}, 29(4):1165--1188.

\bibitem[\protect\astroncite{Brent}{2013}]{brent_algorithms_2013}
Brent, R.~P. (2013).
\newblock {\em Algorithms for {Minimization} {Without} {Derivatives}}.
\newblock Courier Corporation.

\bibitem[\protect\astroncite{Burge}{1988}]{burge_fungi_1988}
Burge, M.~N. (1988).
\newblock {\em Fungi in biological control systems}.
\newblock Manchester University Press.

\bibitem[\protect\astroncite{Cai et~al.}{2019}]{cai_differential_2019}
Cai, T.~T., Li, H., Ma, J., and Xia, Y. (2019).
\newblock Differential {Markov} random field analysis with an application to
  detecting differential microbial community networks.
\newblock {\em Biometrika}, 106(2):401--416.

\bibitem[\protect\astroncite{Carr et~al.}{2019}]{carr_use_2019}
Carr, A., Diener, C., Baliga, N.~S., and Gibbons, S.~M. (2019).
\newblock Use and abuse of correlation analyses in microbial ecology.
\newblock {\em The ISME Journal}, 13(11):2647--2655.

\bibitem[\protect\astroncite{Chen and Li}{2016}]{chen_two-part_2016}
Chen, E.~Z. and Li, H. (2016).
\newblock A two-part mixed-effects model for analyzing longitudinal microbiome
  compositional data.
\newblock {\em Bioinformatics}, 32(17):2611--2617.

\bibitem[\protect\astroncite{Faust et~al.}{2012}]{faust_microbial_2012}
Faust, K., Sathirapongsasuti, J.~F., Izard, J., Segata, N., Gevers, D., Raes,
  J., and Huttenhower, C. (2012).
\newblock Microbial {Co}-occurrence {Relationships} in the {Human}
  {Microbiome}.
\newblock {\em PLOS Computational Biology}, 8(7):e1002606.

\bibitem[\protect\astroncite{Fierer and Jackson}{2006}]{fierer_diversity_2006}
Fierer, N. and Jackson, R.~B. (2006).
\newblock The diversity and biogeography of soil bacterial communities.
\newblock {\em Proceedings of the National Academy of Sciences},
  103(3):626--631.

\bibitem[\protect\astroncite{Fisher and Mehta}{2014}]{fisher_identifying_2014}
Fisher, C.~K. and Mehta, P. (2014).
\newblock Identifying keystone species in the human gut microbiome from
  metagenomic timeseries using sparse linear regression.
\newblock {\em PloS one}, 9(7):e102451.

\bibitem[\protect\astroncite{Friedman and Alm}{2012}]{friedman_inferring_2012}
Friedman, J. and Alm, E.~J. (2012).
\newblock Inferring {Correlation} {Networks} from {Genomic} {Survey} {Data}.
\newblock {\em PLOS Computational Biology}, 8(9):e1002687.

\bibitem[\protect\astroncite{Genest et~al.}{1995}]{genest_semiparametric_1995}
Genest, C., Ghoudi, K., and Rivest, L.-P. (1995).
\newblock A semiparametric estimation procedure of dependence parameters in
  multivariate families of distributions.
\newblock {\em Biometrika}, 82(3):543--552.

\bibitem[\protect\astroncite{Gerber}{2014}]{gerber_dynamic_2014}
Gerber, G.~K. (2014).
\newblock The dynamic microbiome.
\newblock {\em FEBS Letters}, 588(22):4131--4139.

\bibitem[\protect\astroncite{Gilbert et~al.}{2014}]{gilbert_earth_2014}
Gilbert, J.~A., Jansson, J.~K., and Knight, R. (2014).
\newblock The {Earth} {Microbiome} project: successes and aspirations.
\newblock {\em BMC Biology}, 12(1):69.

\bibitem[\protect\astroncite{Greenblum
  et~al.}{2012}]{greenblum_metagenomic_2012}
Greenblum, S., Turnbaugh, P.~J., and Borenstein, E. (2012).
\newblock Metagenomic systems biology of the human gut microbiome reveals
  topological shifts associated with obesity and inflammatory bowel disease.
\newblock {\em Proceedings of the National Academy of Sciences},
  109(2):594--599.

\bibitem[\protect\astroncite{Gunawan et~al.}{2020}]{gunawan_mixed_2020}
Gunawan, D., Khaled, M.~A., and Kohn, R. (2020).
\newblock Mixed {Marginal} {Copula} {Modeling}.
\newblock {\em Journal of Business \& Economic Statistics}, 38(1):137--147.

\bibitem[\protect\astroncite{Ho et~al.}{2019}]{ho_metamicrobiomer_2019}
Ho, N.~T., Li, F., Wang, S., and Kuhn, L. (2019).
\newblock {metamicrobiomeR}: an {R} package for analysis of microbiome relative
  abundance data using zero-inflated beta {GAMLSS} and meta-analysis across
  studies using random effects models.
\newblock {\em BMC Bioinformatics}, 20(1):188.

\bibitem[\protect\astroncite{Joe}{2005}]{joe_asymptotic_2005}
Joe, H. (2005).
\newblock Asymptotic efficiency of the two-stage estimation method for
  copula-based models.
\newblock {\em Journal of Multivariate Analysis}, 94(2):401--419.

\bibitem[\protect\astroncite{Joe and Xu}{1996}]{joe_estimation_1996}
Joe, H. and Xu, J.~J. (1996).
\newblock The {Estimation} {Method} of {Inference} {Functions} for {Margins}
  for {Multivariate} {Models}.

\bibitem[\protect\astroncite{Kurtz et~al.}{2015}]{kurtz_sparse_2015}
Kurtz, Z.~D., Müller, C.~L., Miraldi, E.~R., Littman, D.~R., Blaser, M.~J.,
  and Bonneau, R.~A. (2015).
\newblock Sparse and {Compositionally} {Robust} {Inference} of {Microbial}
  {Ecological} {Networks}.
\newblock {\em PLOS Computational Biology}, 11(5):e1004226.

\bibitem[\protect\astroncite{Lederberg and Mccray}{2001}]{lederberg_ome_2001}
Lederberg, J. and Mccray, A.~T. (2001).
\newblock `{Ome} {Sweet} `{Omics}--{A} {Genealogical} {Treasury} of {Words}.
\newblock {\em The Scientist}, 15(7):8--8.

\bibitem[\protect\astroncite{Li}{2015}]{li_microbiome_2015}
Li, H. (2015).
\newblock Microbiome, metagenomics, and high-dimensional compositional data
  analysis.
\newblock {\em Annual Review of Statistics and Its Application}, 2:73--94.

\bibitem[\protect\astroncite{Liang and Self}{1996}]{liang_asymptotic_1996}
Liang, K.-Y. and Self, S.~G. (1996).
\newblock On the {Asymptotic} {Behaviour} of the {Pseudolikelihood} {Ratio}
  {Test} {Statistic}.
\newblock {\em Journal of the Royal Statistical Society: Series B
  (Methodological)}, 58(4):785--796.

\bibitem[\protect\astroncite{Lozupone and Knight}{2007}]{lozupone_global_2007}
Lozupone, C.~A. and Knight, R. (2007).
\newblock Global patterns in bacterial diversity.
\newblock {\em Proceedings of the National Academy of Sciences},
  104(27):11436--11440.

\bibitem[\protect\astroncite{McDonald et~al.}{2018}]{mcdonald_american_2018}
McDonald, D., Hyde, E., Debelius, J.~W., Morton, J.~T., Gonzalez, A.,
  Ackermann, G., Aksenov, A.~A., Behsaz, B., Brennan, C., Chen, Y., et~al.
  (2018).
\newblock American gut: an open platform for citizen science microbiome
  research.
\newblock {\em Msystems}, 3(3):e00031--18.

\bibitem[\protect\astroncite{McMurdie and Holmes}{2014}]{mcmurdie_waste_2014}
McMurdie, P.~J. and Holmes, S. (2014).
\newblock Waste {Not}, {Want} {Not}: {Why} {Rarefying} {Microbiome} {Data} {Is}
  {Inadmissible}.
\newblock {\em PLOS Computational Biology}, 10(4):e1003531.

\bibitem[\protect\astroncite{Ospina and Ferrari}{2012}]{ospina_general_2012}
Ospina, R. and Ferrari, S. L.~P. (2012).
\newblock A general class of zero-or-one inflated beta regression models.
\newblock {\em Computational Statistics \& Data Analysis}, 56(6):1609--1623.

\bibitem[\protect\astroncite{Paulson et~al.}{2013}]{paulson_differential_2013}
Paulson, J.~N., Stine, O.~C., Bravo, H.~C., and Pop, M. (2013).
\newblock Differential abundance analysis for microbial marker-gene surveys.
\newblock {\em Nature Methods}, 10(12):1200--1202.

\bibitem[\protect\astroncite{Pearson}{1897}]{pearson_mathematical_1897}
Pearson, K. (1897).
\newblock Mathematical contributions to the theory of evolution.—{On} a form
  of spurious correlation which may arise when indices are used in the
  measurement of organs.
\newblock {\em Proceedings of the Royal Society of London},
  60(359-367):489--498.

\bibitem[\protect\astroncite{Peng et~al.}{2015}]{peng_zero-inflated_2015}
Peng, X., Li, G., and Liu, Z. (2015).
\newblock Zero-{Inflated} {Beta} {Regression} for {Differential} {Abundance}
  {Analysis} with {Metagenomics} {Data}.
\newblock {\em Journal of Computational Biology}, 23(2):102--110.

\bibitem[\protect\astroncite{Reshef et~al.}{2011}]{reshef_detecting_2011}
Reshef, D.~N., Reshef, Y.~A., Finucane, H.~K., Grossman, S.~R., McVean, G.,
  Turnbaugh, P.~J., Lander, E.~S., Mitzenmacher, M., and Sabeti, P.~C. (2011).
\newblock Detecting {Novel} {Associations} in {Large} {Data} {Sets}.
\newblock {\em Science}, 334(6062):1518--1524.

\bibitem[\protect\astroncite{Scealy and Welsh}{2011}]{scealy_regression_2011}
Scealy, J.~L. and Welsh, A.~H. (2011).
\newblock Regression for compositional data by using distributions defined on
  the hypersphere.
\newblock {\em Journal of the Royal Statistical Society: Series B (Statistical
  Methodology)}, 73(3):351--375.

\bibitem[\protect\astroncite{Scher and Abramson}{2011}]{scher_microbiome_2011}
Scher, J.~U. and Abramson, S.~B. (2011).
\newblock The microbiome and rheumatoid arthritis.
\newblock {\em Nature Reviews Rheumatology}, 7(10):569--578.

\bibitem[\protect\astroncite{Shih and Louis}{1995}]{shih_inferences_1995}
Shih, J.~H. and Louis, T.~A. (1995).
\newblock Inferences on the {Association} {Parameter} in {Copula} {Models} for
  {Bivariate} {Survival} {Data}.
\newblock {\em Biometrics}, 51(4):1384--1399.

\bibitem[\protect\astroncite{Sklar}{1959}]{sklar_fonctions_1959}
Sklar, M. (1959).
\newblock Fonctions de repartition an dimensions et leurs marges.
\newblock {\em Publ. inst. statist. univ. Paris}, 8:229--231.

\bibitem[\protect\astroncite{Taneja}{2014}]{taneja_arthritis_2014}
Taneja, V. (2014).
\newblock Arthritis susceptibility and the gut microbiome.
\newblock {\em FEBS Letters}, 588(22):4244--4249.

\bibitem[\protect\astroncite{Thompson et~al.}{2017}]{thompson_communal_2017}
Thompson, L.~R., Sanders, J.~G., McDonald, D., Amir, A., Ladau, J., Locey,
  K.~J., Prill, R.~J., Tripathi, A., Gibbons, S.~M., Ackermann, G., et~al.
  (2017).
\newblock A communal catalogue reveals earth’s multiscale microbial
  diversity.
\newblock {\em Nature}, 551(7681):457--463.

\bibitem[\protect\astroncite{Turnbaugh et~al.}{2007}]{turnbaugh_human_2007}
Turnbaugh, P.~J., Ley, R.~E., Hamady, M., Fraser-Liggett, C.~M., Knight, R.,
  and Gordon, J.~I. (2007).
\newblock The {Human} {Microbiome} {Project}.
\newblock {\em Nature}, 449(7164):804--810.

\bibitem[\protect\astroncite{White et~al.}{2009}]{white_statistical_2009}
White, J.~R., Nagarajan, N., and Pop, M. (2009).
\newblock Statistical methods for detecting differentially abundant features in
  clinical metagenomic samples.
\newblock {\em PLoS Comput Biol}, 5(4):e1000352.

\bibitem[\protect\astroncite{Xia et~al.}{2011}]{xia_extended_2011}
Xia, L.~C., Steele, J.~A., Cram, J.~A., Cardon, Z.~G., Simmons, S.~L., Vallino,
  J.~J., Fuhrman, J.~A., and Sun, F. (2011).
\newblock Extended local similarity analysis ({eLSA}) of microbial community
  and other time series data with replicates.
\newblock {\em BMC Systems Biology}, 5(2):S15.

\end{thebibliography}

\newpage

\section*{Appendix}
\subsection*{Proof of theorem 1}

\begin{proof}
	For simplicity, we write $\ell(\theta) = \ell(\theta, \boldsymbol{\tilde{\gamma}}_i, \boldsymbol{\tilde{\gamma}}_j)$. 
	By Taylor expansion, we have
	\[
	\ell(\theta_0) = \ell(\tilde{\theta}) + (\theta_0-\tilde{\theta}) \ell'(\tilde{\theta}) + \frac{1}{2} (\theta_0-\tilde{\theta})^2 \ell''(\tilde{\theta}) + \dots
	\]
	
	Since $\ell'(\tilde{\theta}) = 0$, we have
	
	\[
	\Lambda = -2 [\ell(\theta_0) - \ell(\tilde{\theta})] \asymp -(\theta_0-\tilde{\theta})^2 \ell''(\tilde{\theta}) = 
	-\frac{n(\tilde{\theta} - \theta_0)^2}{ v} \frac{\ell''(\tilde{\theta})v}{n}.
	\]
	where $v=$\textbf{V}$_{7,7}$ is the $(7,7)^{th}$ entry of the covariance matrix of $\boldsymbol{\tilde{\eta}}$, which can be calculated as:
	\beq \label{eq:thetaVar}
	v = \mathcal{I}_{\theta\theta}^{-1} + \mathcal{I}_{\theta\theta}^{-2} 
	( \mathcal{I}_{\theta 1} \mathcal{J}_{11}^{-1} \mathcal{I}_{1 \theta} + 
	\mathcal{I}_{\theta 2} \mathcal{J}_{22}^{-1} \mathcal{I}_{2 \theta} + 
	\mathcal{I}_{\theta 1} \mathcal{J}_{11}^{-1} \mathcal{J}_{12} \mathcal{J}_{22}^{-1} \mathcal{I}_{2 \theta} + 
	\mathcal{I}_{\theta 2} \mathcal{J}_{22}^{-1} \mathcal{J}_{21} \mathcal{J}_{11}^{-1} \mathcal{I}_{1 \theta} ).
	\eeq
	
	Note that $\frac{n(\tilde{\theta}-\theta_0)^2}{v} \xrightarrow{D} \chi^2_1$. Thus, it suffices to deal with the ratio $ \frac{-\ell''(\tilde{\theta})v}{n}$. Now since
	
	\[
	- n^{-1} \ell'' = - n^{-1} \frac{\partial^2 \ell}{\partial \theta^2} = - \frac{1}{n} \sum_{l=1}^n \frac{\partial^2\log f(\textbf{X}_l; \boldsymbol{\tilde{\gamma}}_i, \boldsymbol{\tilde{\gamma}}_j, \theta)}{\partial \theta^2},
	\]
	by the Mean Value Theorem and the Law of Large Numbers,
	\[
	- \ell''(\tilde{\theta})/n \xrightarrow{P} - \ell''(\theta)/n \xrightarrow{P} - \mathbb{E}[\ell''(\theta)] = \mathcal{I}_{\theta\theta},
	\] 
	which can be approximated by $\tilde{\mathcal{I}}_{\theta\theta}$ using numerical methods and $v$ can be estimated by a consistent estimator, $\tilde{v}$, such as the jackknife estimate.
	
	We now define the following two-stage LRT statistic:
	\beq
	\Lambda' = (\tilde{v} \tilde{\mathcal{I}}_{\theta\theta})^{-1} \Lambda = \tilde{\omega} \Lambda
	\eeq
	
	The above discussion implies
	\beq
	\Lambda' \to_D \chi^2_1, \quad \text{as $n\to\infty$.}
	\eeq
\end{proof}

\ignore{
\section*{Tables and Figures}

\begin{table}[h]
	\small
	\centering
	\caption{Normalized AGP network summary statistics. Mean ($\pm$ sd) are reported.}
	\begin{tabular}{cccc}
		\hline 
		\textbf{Degree}     & \textbf{Closeness}  & \textbf{Betweenness} & \textbf{Eigenvector}\\ 
		\hline
		0.625 ($\pm$ 0.144) & 0.735 ($\pm$ 0.075) & 0.006 ($\pm$ 0.003)  & 0.726 ($\pm$ 0.180) \\ 
		\hline 
	\end{tabular}
\label{tbl:networkStats}
\end{table}

\begin{table}[h]
	\centering
	\caption{Global measures of the AGP network significantly different from the average of 1000 Erdős–Rényi models ($p<0.001$).}
	\begin{tabular}{lcc}
		\hline 
		& \textbf{Cluster Coefficient} & \textbf{Modularity} \\ 
		\hline
		AGP       & 0.722          &  0.116 \\ 
		Random & 0.625          & 0.056 \\ 
		\hline
	\end{tabular}
\label{tbl:randomCompare}
\end{table}

}

\ignore{

\begin{figure}[!ht]
	\begin{center}
		\includegraphics[width=15cm]{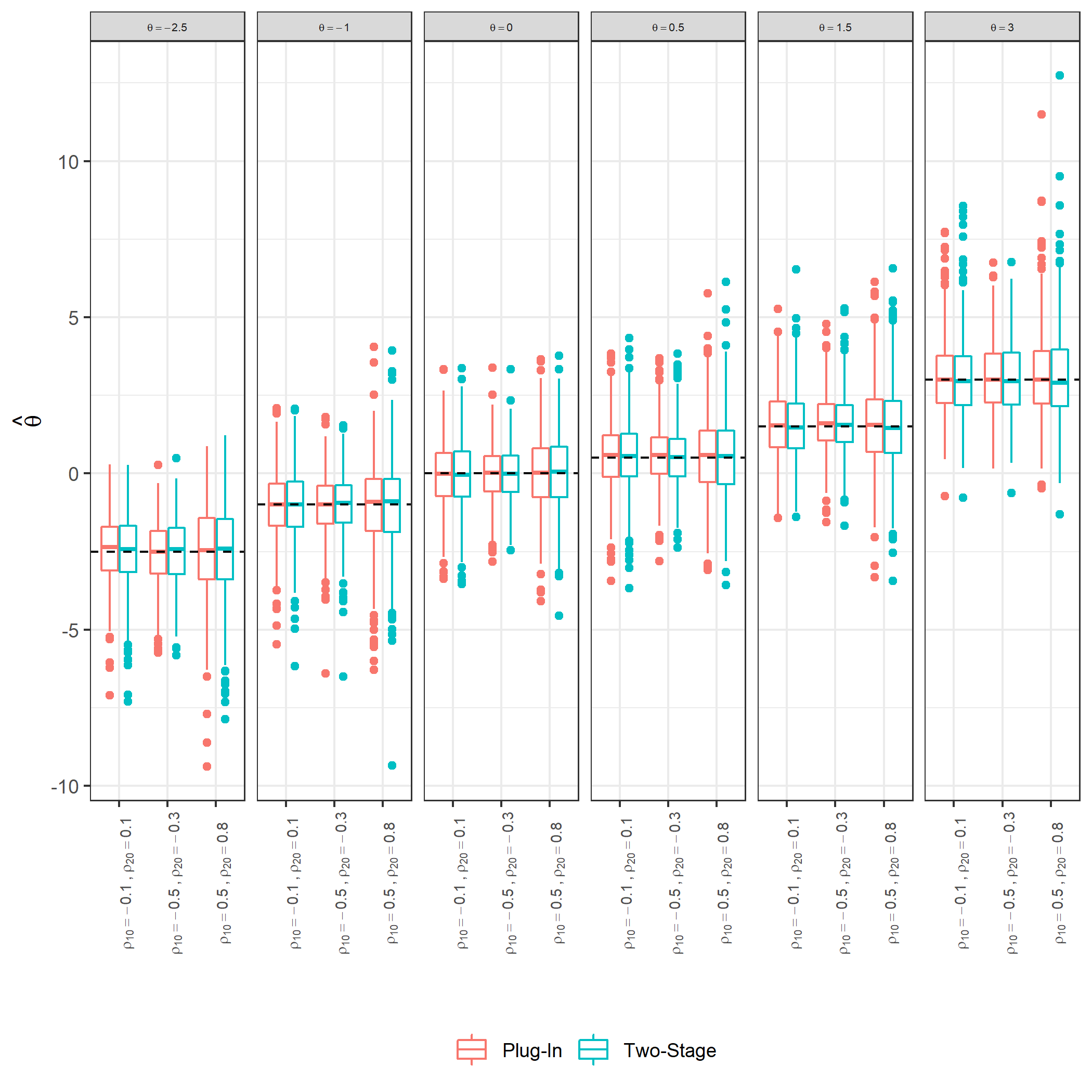}
	\end{center}
	\caption{Boxplot of estimated $\hat\theta$ values across 500 simulations for $n=50$. The black dashed line represents the true $\theta$ value. Data was simulated under varying strength of dependence, mean, dispersion and zero-inflation parameter settings, with one continuous covariate affecting the zero-inflation probabilities.}
	\label{fig:consistency1Q}
\end{figure}
}

\ignore{
\begin{figure}[!ht]
	\begin{center}
		\includegraphics[width=15cm]{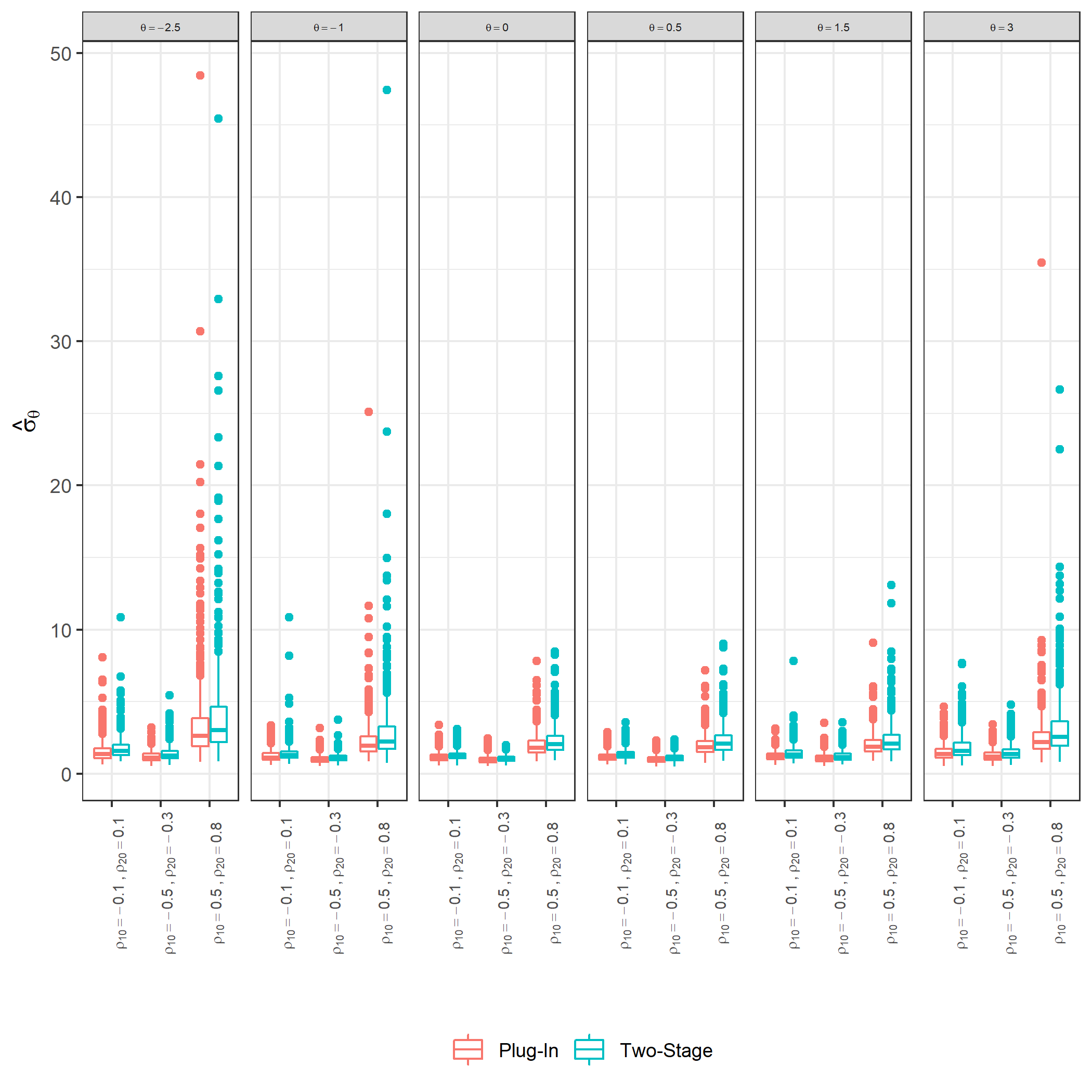}
	\end{center}
	\caption{Boxplot of estimated jackknife variance of $\theta$, denoted as $\hat\sigma_{\theta}^{2}$, across 500 simulations for $n=50$. Data was simulated under varying strength of dependence, mean, dispersion and zero-inflation parameter settings, with one continuous covariate affecting the zero-inflation probabilities. Outliers with variance values greater than 50 were removed from plots.}  
	\label{fig:var1Q}
\end{figure}
}

\ignore{

\begin{figure}[!ht]
	\begin{center}
		\includegraphics[width=15cm]{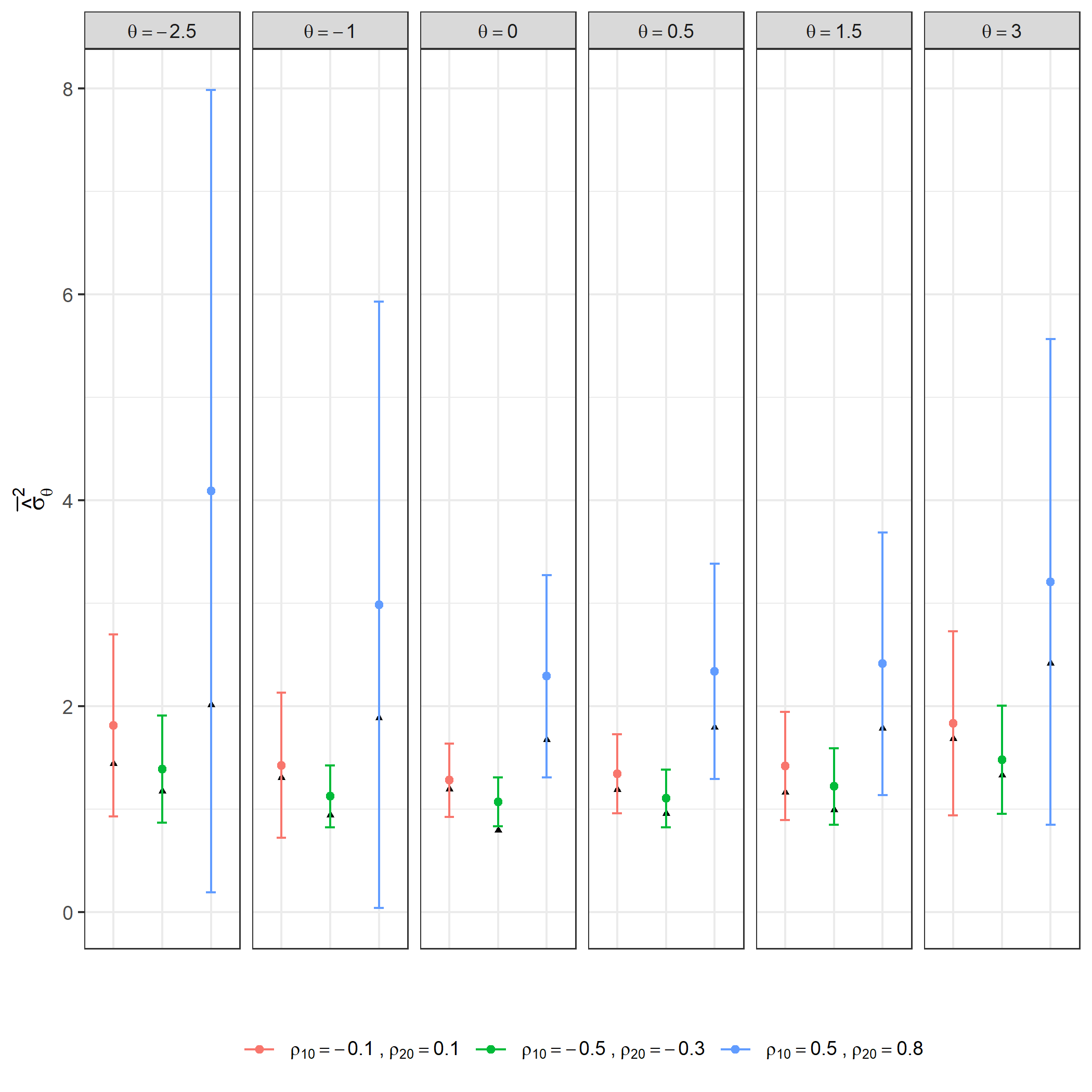}
	\end{center}
	\caption{Mean and standard error (bars) of the estimated jackknife variance of $\tilde\theta$ from data simulated under $n=50$ and varying strength of dependence, mean, dispersion and zero-inflation parameter settings, with one continuous covariate affecting the zero-inflation probabilities. Black triangles correspond the empirical (sample) variance.}
	\label{fig:varBias1Q}
\end{figure}

\begin{figure}[!ht]
	\begin{center}
		\includegraphics[width=15cm]{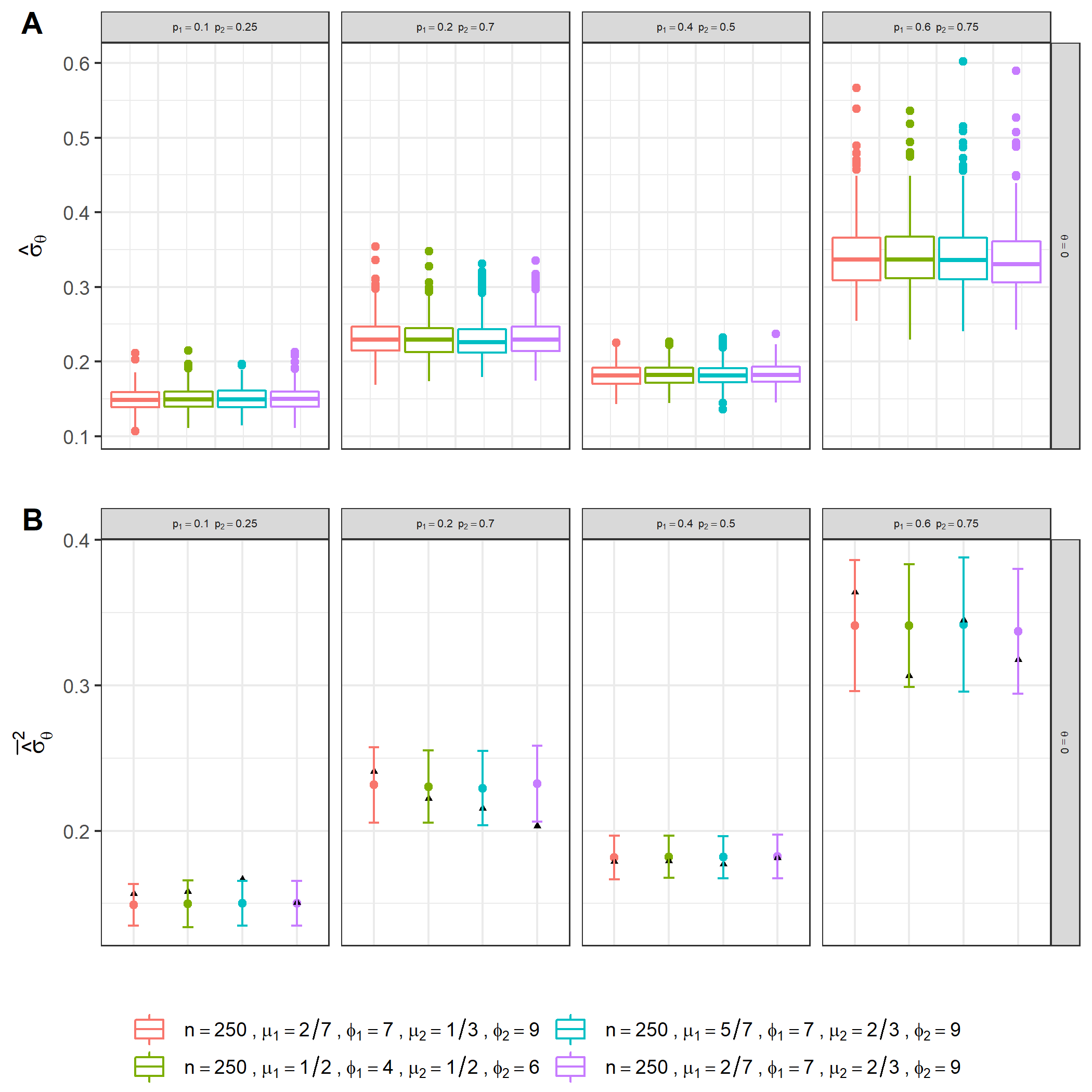}
	\end{center}
	\caption{(A) Boxplot of estimated jackknife variance of $\theta$, denoted as $\hat\sigma_{\theta}^{2}$, across 500 simulations. (B) Mean and standard error (bars) of the estimated jackknife variance of $\tilde\theta$. Black triangles correspond the empirical (sample) variance. Data was simulated under $n=250$ and varying strength of dependence, mean, dispersion and zero-inflation parameter settings, without covariate adjustment.}
	\label{fig:varNC250}
\end{figure}
}

\ignore{
\begin{figure}[!ht]
	\begin{center}
		\includegraphics[width=15cm]{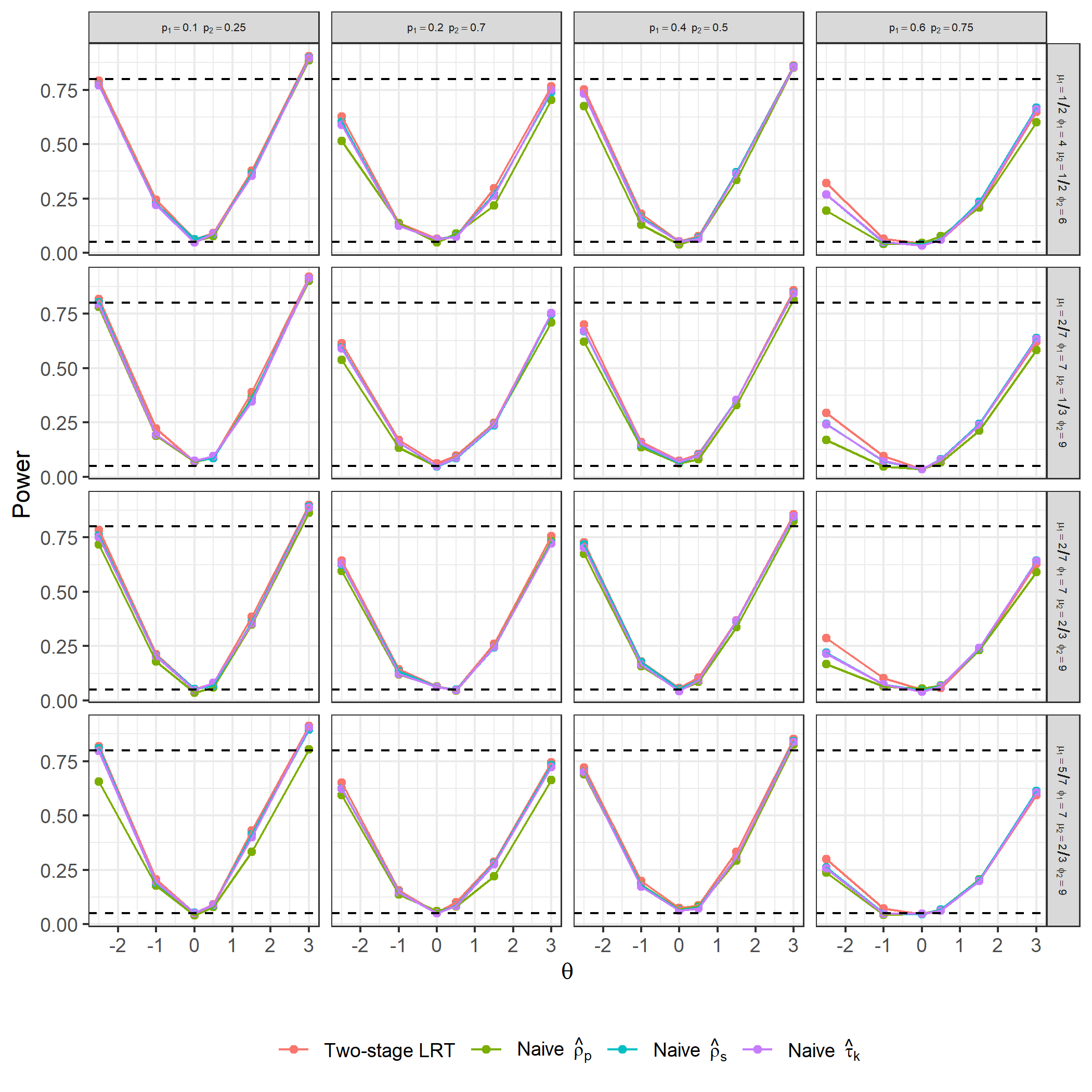}
	\end{center}
	\caption{Power curves of the likelihood ratio test for independence using the Frank copula model with zero-inflated beta margins and a two-stage maximum likelihood estimation procedure. Black dashed lines represent 80\% power and 5\% Type I error. Power was calculated under $n=50$ and varying strength of dependence, mean abundance, dispersion, and zero-inflation probabilities without adjusting for covariates.}
	\label{fig:powerNoCovar}
\end{figure}

\begin{figure}[!ht]
	\begin{center}
		\includegraphics[width=15cm]{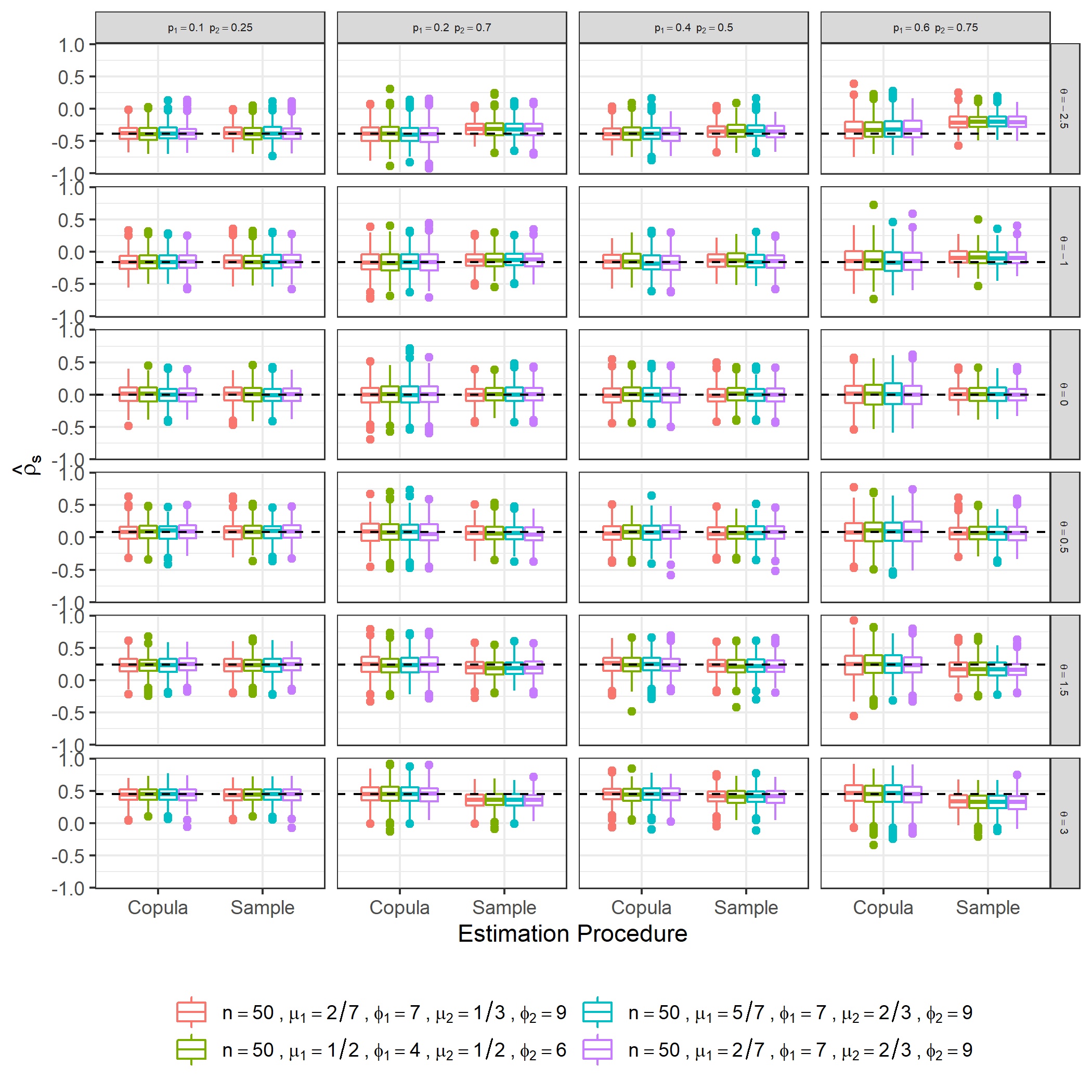}
	\end{center}
	\caption{Boxplot of comparing rank-based and copula $\hat\rho_{s}$ values across 500 simulations. The black dashed line represents the true $\rho_{s}$ value. Data was simulated under varying strength of dependence, mean, dispersion and zero-inflation parameter settings, without covariate adjustment.}
	\label{fig:spCorCompareBP}
\end{figure}
}

\ignore{

\begin{figure}[!ht]
	\centering
	\includegraphics[width=15cm]{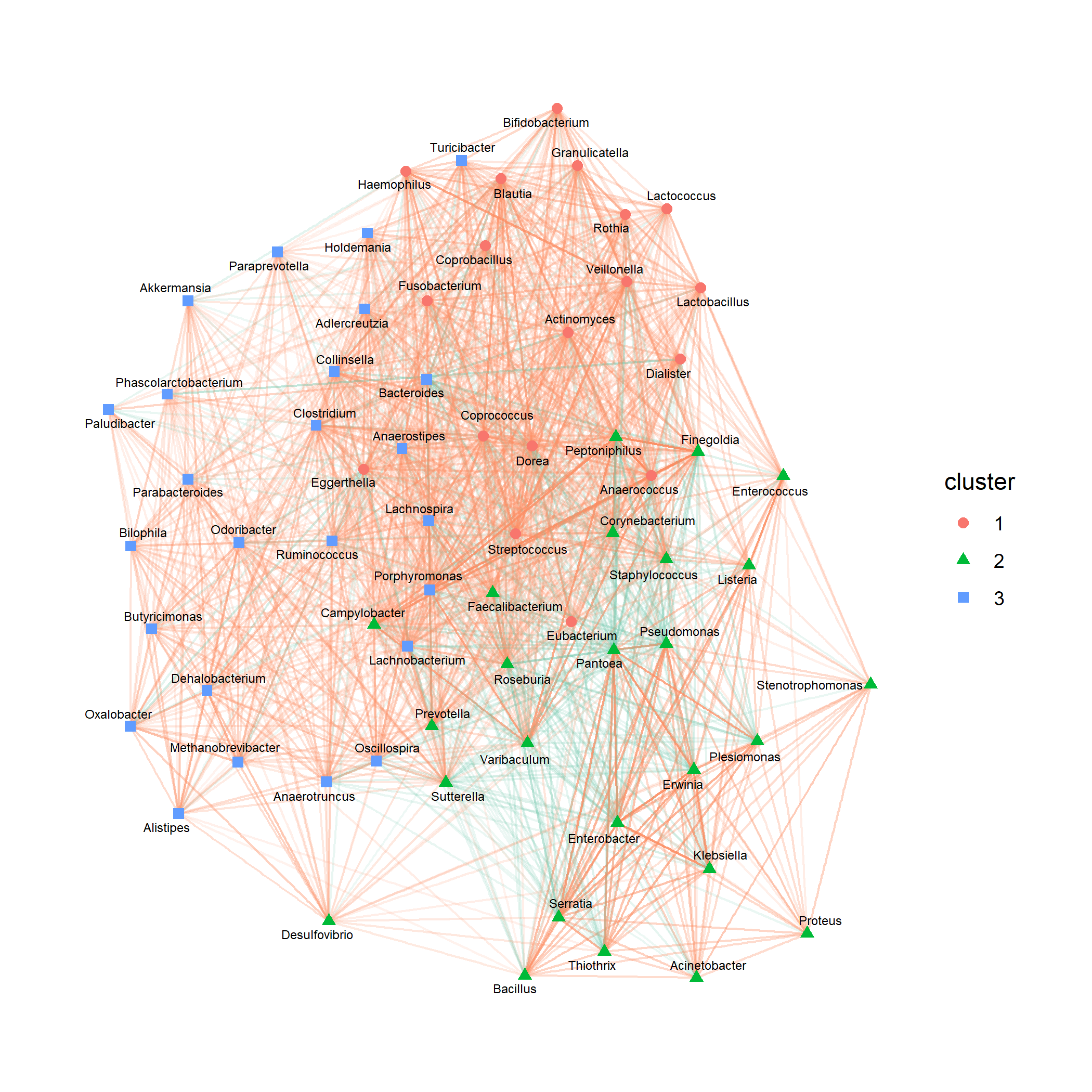}
	\caption{Network graph of the AGP adjacency matrix. Edge color denotes dependency sign with orange edges representing positive (+) theta values and green edges representing negative (-) theta values. Node color and shape represents cluster assignment from a fast greedy algorithm.}
	\label{fig:agpNetwork}
\end{figure}

\begin{figure}[!ht]
	\centering
	\includegraphics[width=15cm]{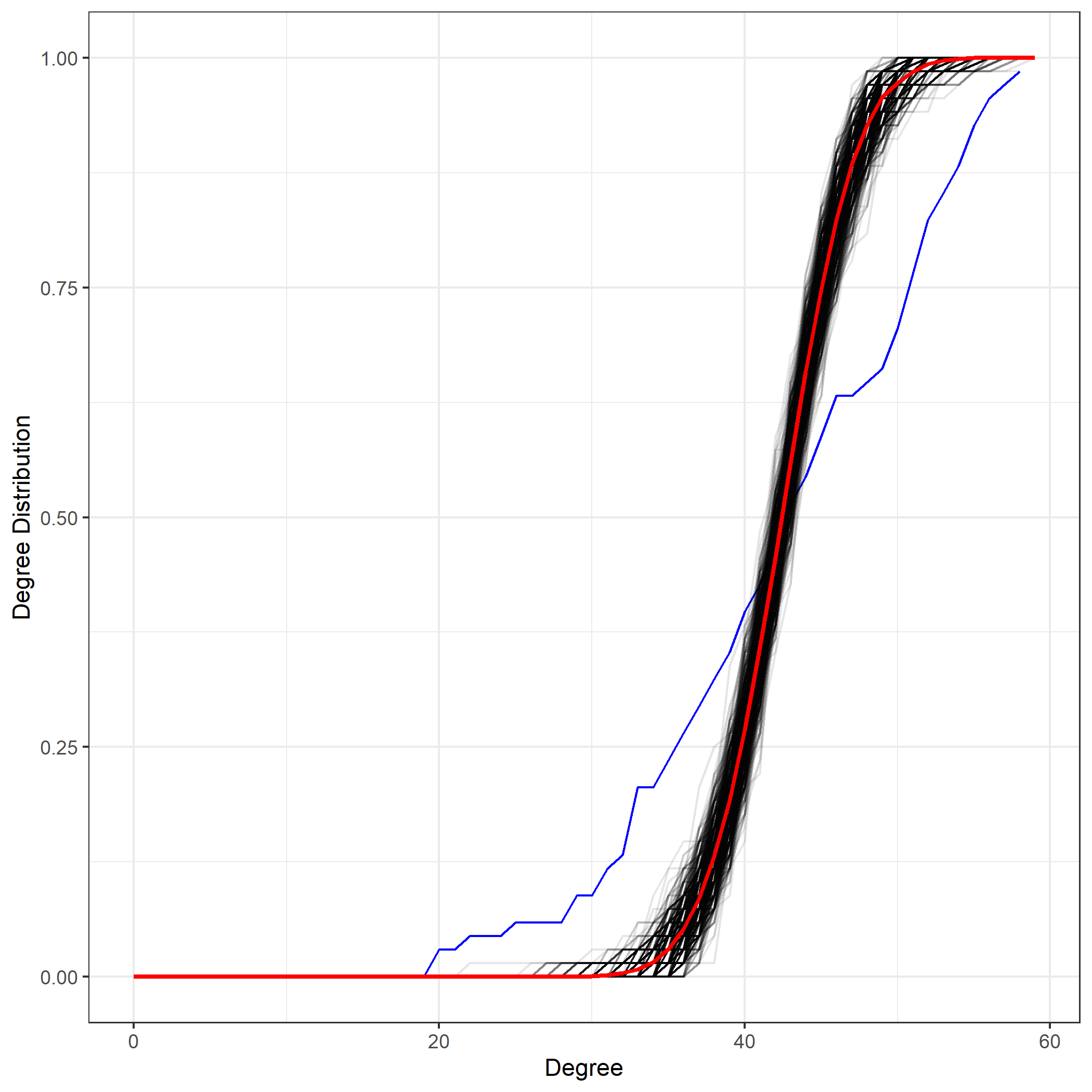}
	\caption{The degree distribution (KS test, $p=0.057$) of the AGP network (blue), 1000 Erdős–Rényi models (black) and their average (red).}
	\label{fig:degreeDist}
\end{figure}

}

\end{document}